\documentclass[final,3p,times,twocolumn,fleqn]{elsarticle}

\usepackage{epstopdf}
\usepackage{amsmath}

\usepackage{amssymb}

\journal{Computers \& Fluids}

\begin{document}

\begin{frontmatter}

\title{Lattice Boltzmann method for semiclassical fluids}


\author[label1,label2]{Rodrigo C. V. Coelho}
\ead{rcvcoelho@if.ufrj.br}

\author[label1,label3]{Mauro M. Doria}
\ead{mmd@if.ufrj.br}
\address[label1]{Departamento de F\'{\i}sica dos S\'{o}lidos, Universidade Federal do Rio de Janeiro, 21941-972 Rio de Janeiro, Brazil}%
\address[label2]{ ETH Z\"{u}rich, Computational Physics for Engineering Materials, Institute for Building Materials, Schafmattstrasse 6, HIF, CH-8093 Z\"{u}rich, Switzerland}
\address[label3]{ Instituto de F\'{\i}sica ``Gleg Wataghin'', Universidade Estadual de Campinas, Unicamp 13083-970, Campinas, S\~ao Paulo, Brazil}

\begin{abstract}
We determine properties of the lattice Boltzmann method for semiclassical fluids, which is based on the Boltzmann equation and the equilibrium distribution function is given either by the Bose-Einstein or the Fermi-Dirac ones. New D-dimensional polynomials, that generalize the Hermite ones, are introduced and we find that the weight that renders the polynomials orthonormal has to be approximately equal, or equal, to the equilibrium distribution function itself for an efficient numerical implementation of the lattice Boltzmann method.
In light of the new polynomials we discuss the convergence of the series expansion of the equilibrium distribution function and the obtainment of the hydrodynamic equations. A discrete quadrature is proposed and some discrete lattices in one, two and three dimensions associated to weight functions other than the Hermite weight are obtained.
We derive the forcing term for the LBM, given by the Lorentz force, which dependents on the microscopic velocity, since the bosonic and fermionic particles can be charged. Motivated by the recent experimental observations of the hydrodynamic regime of electrons in graphene, we build an isothermal lattice Boltzmann method for electrons in metals in two and three dimensions. This model is validated by means of the Riemann problem and of the Poiseuille flow. As expected for electron in metals, the Ohm's law is recovered for a system analogous to a porous medium.
\end{abstract}
%
%


\end{frontmatter}



\section{Introduction}\label{introduction-sec}
The Boltzmann equation~\cite{kremer10} with a Bhatnagar, Gross and Krook (BGK) collision term together with the Maxwell-Boltzmann (MB) equilibrium distribution function (EDF) has found widespread use to describe the flow of classical particles.
This is because it allows for an efficient numerical implementation, which is well fitted to perform simulations in complex geometries such as in porous media~\cite{coelho16}. This is the lattice Boltzmann method (LBM)~\cite{kruger16,succi01}, based on a discretization of the phase space. The equilibrium distribution function is expanded in powers of the macroscopic velocity, under the assumption of a low Mach number, and yet, the hydrodynamical equations remain fully satisfied. This remarkable property follows from an underlying mathematical structure provided by the D-dimensional Hermite polynomials. In this paper, we generalize the present framework to the semiclassical fluids whose constituents are either bosons or fermions.\\

In 1900, Paul Drude explained the transport properties of electrons in materials by treating them like atoms in a rarefied gas whose microscopic velocities satisfy the MB EDF~\cite{ibach03, ashcroft76}. However, electrons are fermions whose microscopic velocities are distributed according to the Fermi-Dirac (FD) distribution instead of the MB one.
Indeed, in 1927, Arnold Sommerfeld  showed that even at room temperature the quantum mechanical properties of the electron gas are relevant.
The gas is essentially governed by their zero temperature properties, where they are piled in energy according to the Pauli exclusion principle.
Only the electrons with the highest energy (Fermi energy) are available for conduction.
Thus, the conduction electrons move with the Fermi speed instead of the thermal microscopic velocity.
Interestingly the Boltzmann-BGK equation provides the standard framework to understand the Drude-Sommerfeld model that describes the electrons in metals~\cite{ibach03}.
In the 1930's E. A. Uheling and G. E. Uhlenbeck~\cite{uehling33} were the first ones to generalize the Boltzmann equation to account for particles obeying either Bose-Einstein (BE) and FD statistics.
In the 1950's Bhatnagar, Gross and Krook proposed to describe collisions among particles in the Boltzmann equation through relaxation of the distribution function to an EDF within a typical time $\tau$. This collision term became fundamental for the development of computationally efficient algorithms to solve the Boltzmann equation.
Nevertheless, only in the 1980's this goal was fully reached by the development of the LBM, which solution of the Boltzmann-BGK equation relies on a discretization of space and time.
The ability to simulate flows in complex geometries had been finally reached and, since then, it has been extensively used to tackle many problems of classical fluid dynamics ranging from biology to material science~\cite{kruger16}.
It features an elegant solution for the quadrature problem, by means of the D-dimensional Hermite polynomials, which is the exact calculation of an integral in a discrete lattice.  The expanded EDF in terms of the ratio between the macroscopic velocity and a reference velocity (Mach velocity) still respects the conservation laws of hydrodynamics~\cite{landau86} in the time evolution process.
However the proposal of such a LBM for semiclassical fluids remained as an open problem, although the existing interest to solve the Boltzmann equation in arbitrary geometries and in presence of granular non-conducting grains (defects or impurities),
Recently, Coelho, Ilha and Doria have proposed a semiclassical LBM to reach this goal, based on  new D-dimensional polynomials~\cite{coelho16-2}. Here, we determine several of its properties such as the choice of the polynomial weight, the Chapman-Enskog derivation of the hydrodynamic equations and the quadrature. We calculate the  expansion up to fourth order of a generic distribution function in D-dimensional polynomials, which is enough to recover up to the energy conservation equation~\cite{coelho14}.
\\

The three fundamental EDFs of statistical mechanics are the Maxwell-Boltzmann (MB) for classical distinguishable particles,  the Fermi-Dirac (FD) for fermions and the Bose-Einstein (BE) for bosons, the last two ones are indistinguishable semiclassical particles.
The quantum statistics takes into account that particles have an intrinsic wavelength, which if larger than their average separation, makes them overlap and turn them indistinguishable from each other. Oppositely in case of low densities the BE-FD statistics reduce to the MB statistics where particles are distinguishable since this overlap is sufficently small to be neglected.
However for a large range of density and temperature quantum effects are still present and for this reason the BE-FD statistics are known as semiclassical statistics whereas MB is a classical statistics. \\

Recently there has been a renewed interest in the study of the hydrodynamic regime for charge carriers in conductors~\cite{narozhny2017hydrodynamic,PhysRevB.93.155122, Lucas23082016, Moll1061}.
Experiments have shown that electrons in graphene exhibit hydrodynamic behavior for a wide range of temperatures and carrier densities~\cite{bandurin16}, due to its weak electron-phonon scattering~\cite{PhysRevLett.103.226801} and to the new technologies to produce ultra-clean samples~\cite{skakalova2014graphene}. One of the clear signals of its hydrodynamical regime is the onset of whirlpools (vortices) that has been predicted and subsequently observed~\cite{PhysRevB.92.165433, pellegrino2016electron, levitov16, bandurin16}. The Dirac fluid of electron  has been simulated by a relativistic LBM many times~\cite{coelho2017kelvin, mendoza11,oettinger13, furtmaier15, mendoza13, giordanelli2017modelling} in order to unveil new properties of graphene. These authors expand the FD distribution in orthogonal polynomials, similarly as done here, but using a fully relativistic formalism for massless particles (see Ref.~\cite{coelho2017kelvin}).
In this paper we seek a general non-relativistic formalism for semiclassical fluids based on D-dimensional orthonormal polynomials. The Hermite polynomials are well fitted to describe classical particles, that is, those obeying the MB statistics, since they are orthonormal under the Hermite weight function which is essentially the MB EDF. Previous attempts~\cite{coelho14, Shi08, yang10, yang09} to build a semiclassical LBM were based on the expansion of the BE-FD distributions in Hermite polynomials, but they were limited to a nearly classical regime since the weight function of the Hermite polynomials differs greatly from the BE-FD distributions on the semiclassical or quantum regimes (low temperatures and/or high densities).
The point of view here is that a new polynomial set must be used for semiclassical fluids where the weight function is similar or equal to the EDF. In this paper, we propose the general formalism to describe the flow of semiclassical particles based on a new set of D-dimensional polynomials that generalize the well-known D-dimensional Hermite polynomials. We obtain the expansion of the EDF under a general weight such that the cases of BE-FD EDFs can be immediately treated by the present formalism. \\

The description of electrons in metals with Boltzmann equation meets the following parameters~\cite{ashcroft76}.
The microscopic velocity of electrons is the Fermi speed, $ v_F \sim 10^6\, $ m/s and  the typical relaxation time is $\tau \sim 10^{-14}\,$ s. This renders a  kinematic viscosity $\nu \approx v_F^2\tau/3 \sim 10^{-3}\,$ m$^{2}$/s.
The macroscopic velocity $u$ is very low  and can be computed by assuming that $u \, m_e/\tau = e E$, where $m_e$ and $e$ are the electron's mass and charge respectively.
For a typical home appliance battery, the voltage is $V=1.5\,$ Volts per $L=4$ cm, which gives an electric field of $E=V/L \sim 0.4\times10^{2}\,$ Volts/m and so, $u \sim 0.4\,$ m/s. The  typical electronic density is $\rho \sim 10^{-2}\,$ Kg/m$^{3}$. and the shear viscosity is $\eta = \rho \nu \sim 10^{-5}\,$ Kg/(ms).
Thus one can get an estimative for the Reynolds number associated to a system of size $L$, $R=uL/\nu \sim 4.0 \times
10^{-4} L\,$ where $L$ is expressed in meters. Therefore electrons in metals behave similarly to Glycerin. \\

This paper is organized as follows. In Sec. \ref{edf-expansion-sec} we review the expansion of a EDF in terms of orthogonal polynomials. We also discuss the special case of a weight function equal to the EDF itself.
In Sec.~\ref{polynomials-sec} we show the generalized polynomials.
In Sec.~\ref{general-weight} the EDF is expanded up to fourth order in the set of new generalized polynomials orthonormal under a general weight.
In Sec.~\ref{special-weight} the EDF is expanded in polynomials orthonormal for the special case that the weight is equal to the EDF itself, $\omega(\xi)=f^{eq}(\xi)$.
In Subsec.~\ref{edfn2} we directly derive the EDF to order N=2 without using the orthonormal polynomials just to confirm the rightness of our ideas.
In Sec. \ref{hydrodynamic-sec}, we derive the macroscopic equations for semiclassical fluids (i.e., mass, momentum and energy conservation equations) in the context of the generalized polynomials. A discussion is made about the minimum order that the EDF should be expanded in order to recover each macroscopic equation. In Sec.~\ref{forcing-sec}, we calculate the forcing term for a second order expansion of the semiclassical EDF and verify that it satisfies the moment constraints up to second order. In Sec.~\ref{quadrature-sec}, we obtain the quadratures and calculate the discrete weights of the $D1V3$, $D2V9$, and $D3V15$ lattices (more quadratures can be found in \ref{quadrature-appendix}). They are calculated for a generic weight function similarly to the polynomials. In Sec.~\ref{lbm-electrons-sec}, we develop our LBM for electrons in metals in two (2D) and three (3D) dimensions and perform three numerical tests: the Riemann problem, the Poiseuille flow and the Ohm's law. In Sec.~\ref{conclusion-sec}, we summarize our main results and conclude.

\section{Expansion of the equilibrium distribution function}\label{edf-expansion-sec}
The equilibrium distribution function (EDF) $f^{(eq)}(\boldsymbol{\xi})$  is a central quantity in the Boltzmann-BGK framework since the non-equilibrium distribution function $f$ relaxes to the EDF, $f^{eq}$, within time $\tau$, according to,
\begin{flalign} \label{boltz-cont}
\frac{\partial f}{\partial t}+ \boldsymbol{\xi}\cdot \boldsymbol{\nabla}_{\boldsymbol{x}} f + \boldsymbol{a}\cdot \boldsymbol{\nabla}_{\boldsymbol{\xi}} f= - \frac{f-f^{eq}}{\tau}.
\end{flalign}
where $\boldsymbol{\xi}$ and $\boldsymbol{a}$ are the microscopic velocity and acceleration, the latter defined by the external applied force.
From this equation one obtains the hydrodynamical quantities under the  Chapman-Enskog assumption, which says that expectation values over the microscopic velocity, $\boldsymbol{\xi}$,
can be computed either from the non-equilibrium distribution function, $f(\boldsymbol{\xi})$, or from the equilibrium one, $f^{(eq)}(\boldsymbol{\xi})$. In this case the first three macroscopic moments  are given by,
\begin{flalign}
&\rho \equiv \int d^D \boldsymbol{\xi}f^{(eq)}(\boldsymbol{\xi}-\boldsymbol{u}),\label{rho-def-eq}\\
&\rho \boldsymbol{u}\equiv \int d^D \boldsymbol{\xi}f^{(eq)}(\boldsymbol{\xi}-\boldsymbol{u})\boldsymbol{\xi}, \label{vel-def-eq}\\
&\rho \bar\theta \delta_{i_1i_2} \equiv \int d^D \boldsymbol{\xi} f^{(eq)}(\boldsymbol{\xi}-\boldsymbol{u}) (\xi-u)_{i_1}(\xi-u)_{i_2}.\label{theta-bar-def-eq}
\end{flalign}
They correspond to the mass density $\rho$, the macroscopic velocity $\boldsymbol{u}$, and the temperature related quantity $\bar \theta$, which gives a measure of the energy density $\varepsilon=D\bar\theta /2$ (for the classical case, it is the temperature itself: $\bar\theta = \theta$).
Therefore the use of $f(\boldsymbol{\xi}-\boldsymbol{u})$ instead of
$f^{(eq)}(\boldsymbol{\xi}-\boldsymbol{u})$ renders the same values at any position and time for $\rho$, $\boldsymbol{u}$, and  $\bar \theta$ according to the Chapman-Enskog assumption.
We also define for later purposes the quantity $g$ associate to the fourth order expectation value,
\begin{flalign}
& \rho {\bar\theta}^2 g \,\delta_{i_1i_2i_3i_4}\equiv \int d^D \boldsymbol{\xi}f^{(eq)}(\boldsymbol{\xi}-\boldsymbol{u}) \nonumber \\
&\cdot (\xi-u)_{i_1}(\xi-u)_{i_2}(\xi-u)_{i_3}(\xi-u)_{i_4}, \label{gz-def-eq}
\end{flalign}
which is a function of the same parameters that define $\rho$ and $\bar \theta$.
Notice that the definition of the macroscopic velocity $\boldsymbol{u}$ is just the statement that at the local center of mass there is no net motion.
\begin{flalign*}
\int d^D \boldsymbol{\xi}f^{(eq)}(\boldsymbol{\xi}-\boldsymbol{u})(\boldsymbol{\xi}-\boldsymbol{u})=0
\end{flalign*}
The D-dimensional Euclidean space is endowed with the following tensors, defined by Harold Grad~\cite{grad49}, that can be expressed as sums of products of the Kronecker's delta function ($\delta_{ij}=1$ for $i=j$ and 0 for $i\neq j$),
\begin{flalign}
&\delta_{i_1\cdots i_N\vert j_1\cdots j_N} \equiv \delta_{i_1 j_1}\cdots \delta_{i_Nj_N}\,+\,\mbox{permutations  of  $i$'s},
\label{delta0}\\ &
\delta_{i_1\cdots i_N\, j_1\cdots j_N} \equiv \delta_{i_1 j_1}\cdots \delta_{i_Nj_N}\,+\, \mbox{ all permutations}.
\label{delta1}
\end{flalign}
Properties of the above tensors, such as their number of terms, is  discussed in more details  Ref.~\cite{doria17}.

Using the Chapman-Enskog expansion~\cite{chapman70, kremer10}, the macroscopic equations and the transport coefficients for a fluid governed by a given EDF $f^{eq}$ can be calculated from the discrete Boltzmann equation,
\begin{flalign}\label{boltz-disc-eq}
&f(\boldsymbol{x}+ \boldsymbol{\xi}\Delta t, \boldsymbol{\xi}+\boldsymbol{a}\Delta t, t + \Delta t) \nonumber \\ &
- f(\boldsymbol{x}, \boldsymbol{\xi}, t)
= -\Delta t\,\frac{f-f^{eq}}{\tau},
\end{flalign}
where $\Delta t$ is the step time.
The continuous form of this equation is obtained by expanding in powers of $\Delta t$ and the first order term is the continuous Boltzmann equation, given Eq.\eqref{boltz-cont}.\\

We are interested in expanding the EDF in a polynomial power series in  D-dimensional space, as such,
\begin{flalign}
f^{(eq)}(\boldsymbol{\xi} -\boldsymbol{u})= \omega( \boldsymbol{\xi} ) \sum_{N=0}^{K}\frac{1}{N!} \mathcal{A}_{i_1\,i_2 \cdots i_N}(\boldsymbol{u} ) \mathcal{P}_{i_1\,i_2\cdots i_N}( \boldsymbol{\xi}).
\label{feq-complete-exp-eq}
\end{flalign}
$\mathcal{A}_{i_1\,i_2 \cdots i_N}$ are the projections of the EDF on the polynomials and $\mathcal{P}_{i_1\cdots i_N}( \boldsymbol{\xi})$ are the polynomials themselves, which are orthonormal under a generic weight function $\omega(\boldsymbol{\xi} )$. This weight function is assumed to only depend on the modulus of the microscopic velocity $\boldsymbol{\xi} \equiv (\xi_1,\xi_2,\cdots,\xi_D)$: $\omega(\boldsymbol{\xi} )= \omega(\xi )$, $\xi \equiv\vert \boldsymbol{ \xi} \vert$, and to have the property that $\omega(\boldsymbol{\xi}) \rightarrow 0$ for $ \xi \rightarrow \infty$ faster than any power of $\xi$.\\

Remarkably only a few terms in the above series expansion must be included to guarantee mass, momentum and energy conservation. The Chapman-Enskog analysis shows that in order to obtain the hydrodynamic equations of continuity and  the Navier-Stokes equation the above series expansion can be cutoff at K=3 and to include the equation of energy conservation one must include only one more term, namely, go to order K=4.
More details about the obtainment of the macroscopic equations can be found in Sec.~\ref{hydrodynamic-sec}.
However the above cutoff procedure does not guarantee convergence of the  cutoff series expansion to the orginal EDF.
Hence it remains another very important requirement to be fullfilled which is of convergence to the underlying physics. This is expected under the assumption that the macroscopic velocity is much smaller than the microscopic one,  $u  \ll \xi$.
The point that we stress here is that clearly this convergence is intimately connected to the choice of the weight function $\omega(\xi )$, that must be  close enough to $f^{(eq)}(\xi)$ such that the remaining  multiplying  series is just a small correction to it and so, only a few terms would be enough to describe the corrections in $\boldsymbol{\xi}$.\\

The three  EDFs explicitly depend on the microscopic velocity, $\boldsymbol{\xi}$, and implicitly on the position, $\boldsymbol{x}$, through the local macroscopic parameters, such as the density $\rho(\boldsymbol{x})$, the chemical potential $\mu(\boldsymbol{x})$, the fugacity $z(\boldsymbol{x})$ the macroscopic velocity $\boldsymbol{u}(\boldsymbol{x})$, and the temperature $\theta(\boldsymbol{x})$. The FD ($+$) and the BE ($-$) EDFs  are given by
\begin{flalign}\label{be-fd-dist-eq}
 f^{(eq)}_{\,FD/BE}(\boldsymbol{\xi}) = \frac{1}{z^{-1}\exp\left (\boldsymbol{\xi}^2/2 \theta\right)\pm 1}, \; z =\exp{\left(\mu/\theta\right)}
 \end{flalign}
and the MB is,
\begin{flalign}
f^{(eq)}_{MB}(\boldsymbol{\xi}) = \frac{\rho_0}{(2 \pi \theta )^{D/2}} \exp\left(-\boldsymbol{\xi}^2/2 \theta\right )
\end{flalign}
where $\rho_0$ is a dimensionless density. All variables are defined dimensionless $(m=k_B=c=\hbar=e=1)$ by means of appropriate temperature and velocity scales.
We observe the remarkable property of the series expansion in terms of the new polynomials, which is to have the macroscopic velocity expressed always as a ratio to a reference velocity of the problem, not necessarily the thermal velocity. For instance, notice that, in case of electron's in metals, Mach's number diverges at $T=0$ and cannot even be considered as a reference velocity. Indeed in this case the scale is set by the Fermi speed, $v_F$ and, since the microscopic velocity $U$ is very low in metals, for typical electric fields,  $u/v_F \sim 0.4 \times 10^{-6}$, according to previously given values. Hence the new polynomials in case of electrons in metals  render the series expansion of the EDF at $T=0$ to be automatically  in powers of the ratio $u/v_F$ instead, as shown in Ref.~\cite{coelho16-2}. \\

The convergence problem is better understood in the limit of a vanishing  macroscopic velocity at Eq.\eqref{feq-complete-exp-eq}, which gives that,
\begin{flalign}
f^{(eq)}(\boldsymbol{\xi})= \omega( \xi ) \sum_{N=0}^{K}\frac{1}{N!} \mathcal{A}_{i_1\,i_2 \cdots i_N}(0) \mathcal{P}_{i_1\,i_2\cdots i_N}( \boldsymbol{\xi}).
\label{feq-uzero}
\end{flalign}
$f^{(eq)}(\boldsymbol{\xi})$ for the MB EDF is a gaussian and so the choice of $\omega(\xi)$ equal to the Hermite weight function,
\begin{eqnarray}\label{hermite-weight-eq}
\omega(\boldsymbol{\xi}) = \frac{1}{(2 \pi )^{D/2}} \exp\left(-\frac{\boldsymbol{\xi}^2}{2} \right ),
\end{eqnarray}
is a good choice which implies the presence of the D-dimensional Hermite polynomials.  However the Hermite weight is very different from the FD or MB EDFs rendering the need of many terms in the series expansion that multiplies $\omega(\xi)$.
A striking example of such difficulty is provided by the FD EDF very near to zero temperature ($\theta \approx 0$) which becomes equal to a Heaviside step function ($f^{(eq)}_{\,FD}\approx 1$ for $\boldsymbol{\xi}^2/2 \le \mu$ and $f^{(eq)}_{\,FD} \approx 0$ for $\boldsymbol{\xi}^2/2 \ge \mu$). Expressing the Heaviside function as the product of a gaussian times a series expansion implies that the latter must contain many terms. Hence this zero limit provides evidence of the inadequacy of the Hermite weight to describe electrons in metals for instance, since those behave at room temperature very similarly to the zero temperature limit~\cite{ashcroft76}. Hence we reach the conclusion that for indistinguishable particles, namely BE or FD particles,  the weight function must be chosen accordingly in order to reach convergence in the expanded EDF with a few terms.
In this paper we study separately two similar but distinguishable situations, namely, of  the weight similar to the EDF itself, $\omega( \xi ) \approx f^{(eq)}(\boldsymbol{\xi})$ and  of the weight exactly equal to the EDF, $\omega( \xi ) =  f^{(eq)}(\boldsymbol{\xi})$. In order to treat these two cases we firstly obtain the D-dimensional polynomials orthonormal under a general weight $\omega( \xi )$.

\section{The generalized polynomials}\label{polynomials-sec}

The orthonormality condition satisfied by the polynomials in this D-dimensional Euclidean space is given by,
\begin{flalign}
&\int d^D \boldsymbol{\xi} \, \omega( \boldsymbol{\xi} )\mathcal{P}_{i_1\cdots i_N}( \boldsymbol{\xi})\mathcal{P}_{j_1\cdots j_M}(\boldsymbol{\xi})=
\delta_{\scriptscriptstyle {N M}}\delta_{i_1\cdots i_N\vert j_1\cdots j_M}.
\label{omeg-feq}
\end{flalign}
The polynomials $\mathcal{P}_{i_1 \cdots i_N}(\boldsymbol{\xi})$ are expressed in terms the  components $\xi_{i}$ and of the Kronecker´s delta function, $\delta_{ij}$.
The  N$^{\mbox{th}}$ order polynomial  is symmetrical in the indices $i_1 \cdots i_N$, and its parity  is $(-1)^N$.
\begin{flalign*}
 &\mathcal{P}_{i_1 \cdots i_N}(-\xi_{i_1},\ldots, -\xi_{i_k},\ldots,-\xi_{i_N})=\nonumber\\&  (-1)^{N}\mathcal{P}_{i_1 \cdots i_N}(\xi_{i_1},\ldots, \xi_{i_k},\ldots \xi_{i_N})
\end{flalign*}
The N$^{th}$ order polynomial is the sum of all possible symmetric tensors built from products of $\xi_{i}$ and of $\delta_{ij}$ times coefficients which are themselves polynomials in $\xi^2$ to maximum allowed power. This recipe yields a unique expression for the N$^{th}$ order polynomial.
As a working example we take the first five (N=0 to 4) polynomials,
\begin{flalign*}
&\mathcal{P}_0(\boldsymbol{\xi}) = c_0,\\
&\mathcal{P}_{i_1}(\boldsymbol{\xi})=c_1\,\xi_{i_1},\\
&\mathcal{P}_{i_1 i_2}(\boldsymbol{\xi})=c_2\,\xi_{i_1} \xi_{i_2} +f_2(\xi)\,\delta_{i_1 i_2}, \nonumber \\ &  \mbox{where} \; f_2(\xi) \equiv {\bar c}_2 \xi^2+ {c^\prime}_2,\\
&\mathcal{P}_{i_1 i_2 i_3}(\boldsymbol{\xi})=c_3 \,\xi_{i_1} \xi_{i_2}\xi_{i_3}  + f_3(\xi)\big (\xi_{i_1} \delta_{i_2 i_3} + \xi_{i_2}\delta_{i_1 i_3} \nonumber \\ &+  \xi_{i_3} \delta_{i_1 i_2}\big), \mbox{where} \; f_3(\xi) \equiv {\bar c}_3 \xi^2+ {c^\prime}_3,\\
&\mathcal{P}_{i_1 i_2 i_3 i_4}(\boldsymbol{\xi})=c_4 \,\xi_{i_1} \xi_{i_2}\xi_{i_3} \xi_{i_4}+  f_4(\xi)\,\big(\xi_{i_1}\xi_{i_2}
\delta_{i_3 i_4} \nonumber \\ &+\xi_{i_1}\xi_{i_3}
\delta_{i_2 i_4} +\xi_{i_1}\xi_{i_4}
\delta_{i_2 i_3} +\xi_{i_2}\xi_{i_3} \delta_{i_1 i_4}+ \xi_{i_2}\xi_{i_4}
\delta_{i_1 i_3} \nonumber \\ &+\xi_{i_3}\xi_{i_4}\delta_{i_1 i_2} \big )  + g_4(\xi)\,\delta_{i_1 i_2 i_3 i_4},\; \mbox{where} \; f_4(\xi) \equiv \nonumber \\ &\big( {\bar c}_4 \xi^2+ {c^\prime}_4 \big),  \mbox{and} \; g_4(\xi) \equiv \big( {\bar d}_4 \xi^4+{d^\prime}_4 \xi^2+  {d}_4 \big).
\end{flalign*}
These first five polynomials sum to a total of 14 coefficients (1 for N=0, 1 for N=1, 3 for N=2, 3 for N=3 and 6 for N=4) to be determined from the orthonormality condition of Eq.(\ref{omeg-feq}).
The explicit orthonormalization procedure of the first five polynomials (N=0 to 4) produces exactly the 14 equations needed to determine them ($c_K$ for $K=0,1,2,3,4$, ${c^\prime}_K$ for $K=2,3,4$, ${\bar c}_K$ for $K=2,3,4$, $d_4$, ${d^\prime}_4$ and ${\bar d}_4$). This remarkable matching between the orthonormality condition and the D-dimensional Euclidean space symmetry group makes us conjecture that the present method of determining the coefficients can be extended to the N$^{th}$ order. There is a deep tensorial structure behind the orthonormal condition of Eq.(\ref{omeg-feq}).
This structure and the  number of terms in tensors defined in Eqs.(\ref{delta0}) and (\ref{delta1}) are  discussed elsewhere in  Ref.~\cite{doria17}.
\\

The coefficients of the polynomials are solely functions of the weight $\omega(\xi)$ through the integrals $I_{N}$, which are central to the present study. They are assumed to exist and to have well defined properties.
\begin{flalign}\label{i2n-cont}
I_{N}\,\delta_{i_1\cdots i_{N}} \equiv \int d^D \boldsymbol{\xi} \, \omega( \boldsymbol{\xi} ) \, \xi_{i_1}\cdots \xi_{i_{N}}
\end{flalign}
By symmetry it holds that $I_{2N+1}=0$ since the integral vanishes.
Using the spherical integration volume, $\int d^D \boldsymbol{\xi} \, \omega( \boldsymbol{\xi} )= D\pi^{D/2}/\Gamma(D/2+1) \int d\xi\,\xi^{D-1} \omega(\xi) $, the $I_{2N}$ integrals become,
\begin{flalign}\label{i2n-cont2}
I_{2N}=\frac{\pi^{\frac{D}{2}}}{2^{N-1}\Gamma\big(N+\frac{D}{2}\big)}\int_{0}^{\infty}d\xi\, \omega(\xi)\,\xi^{2N+D-1}.
\end{flalign}
The  explicitly derivation of the coefficients by imposing the orthonormality of the first five polynomials is carried in Ref.~\cite{doria17}. The coefficients can be summarized as follows:
\begin{flalign*}
& c_K= \frac{1}{\sqrt{I_{2K}}},\:\:\:\: \mbox{for  $K=0,1,2,3,4$}, \nonumber \\
& {c^\prime}_K=-c_K \frac{I_{2K-2}}{I_{2K-4}}\Delta_{2K-2},\:\:\:\: \mbox{for  $K=2,3,4$},  \nonumber \\
& {\bar c}_K=c_K \frac{\left (-1+\Delta_{2K-2}\right)}{D+2K-4},\:\:\:\: \mbox{for  $K=2,3,4$}, \nonumber \\
& \Delta_{2K}= \sqrt{\frac{2}{\big(D+2K\big)-J_{2K}\big(D+2K-2\big)}} \nonumber \\
& J_{2K}= \frac{I_{2K}^2}{I_{2K+2}I_{2K-2}} \nonumber
\end{flalign*}
\begin{flalign*}
&d_4^2=\frac{8\delta_4^2I_4}{\delta_2\left[\delta_2\delta_6\left(D+4\right)-\delta_4^2 D \right]},\\
&d^\prime_4= -\frac{d_4}{D}\left[\frac{I_0}{I_2}+\frac{I_4\delta_2}{I_2\delta_4}\right]+\frac{2c_4I_6\Delta_6}{D I_4}, \\
&{\bar d}_4 = \frac{d_4 \delta_2}{ D\left(D+2\right)\delta_4}+\frac{c_4\left[D-2\left(D+2\right) \Delta_6 \right]}{D\left(D+2\right)\left(D+4\right)},
\end{flalign*}
where $\delta_L \equiv 2I_{L+2}I_{L-2}/\Delta_L^2$.

The well-known D-dimensional Hermite polynomials are just a particular case of the present polynomials for the case of a Hermite weight,
\begin{flalign}\label{weight-func-hermite-poly-eq}
\omega(\xi) = \frac{1}{(2\pi)^{D/2}}\exp\big({-\frac{\xi^2}{2}}\big).
\end{flalign}
To obtain the integrals $I_{2N}$ of Eq.(\ref{i2n-cont2}), we note that for the above $\omega(\xi)$,
\begin{flalign}\label{i2n-cont2-2}
\int_{0}^{\infty} d\xi\, \omega(\xi)\,\xi^{2N+D-1}= \frac{2^{N-1}}{\pi^{\frac{D}{2}}}\Gamma\big(N+\frac{D}{2}\big).
\end{flalign}
Then it follows from Eq.(\ref{i2n-cont2-2}) that,
\begin{flalign}\label{i2n-hermite}
I_{2N}=1.
\end{flalign}
In this limit $c_K=1$, $\bar c_K=0$  $c^{\prime}_K=-1$,  $d_4=1$, $\bar d_4=0$ and $d^{\prime}_4=0$, and the polynomials become,
\begin{flalign*}
&\mathcal{P}_0(\boldsymbol{\xi}) = 1, \:\:\:\mathcal{P}_{i_1}(\boldsymbol{\xi})=\,\xi_{i_1}, \:\:\:\mathcal{P}_{i_1 i_2}(\boldsymbol{\xi})=\,\xi_{i_1} \xi_{i_2} - \delta_{i_1 i_2}\\
&\mathcal{P}_{i_1 i_2 i_3}(\boldsymbol{\xi})= \,\xi_{i_1} \xi_{i_2}\xi_{i_3}  -\big (\xi_{i_1} \delta_{i_2 i_3} + \xi_{i_2}\delta_{i_1 i_3}  +\xi_{i_3} \delta_{i_1 i_2}\big),  \\
&\mathcal{P}_{i_1 i_2 i_3 i_4}(\boldsymbol{\xi})= \,\xi_{i_1} \xi_{i_2}\xi_{i_3} \xi_{i_4}-\,\big(\xi_{i_1}\xi_{i_2}
\delta_{i_3 i_4} +\xi_{i_1}\xi_{i_3}
\delta_{i_2 i_4} \nonumber \\
&+\xi_{i_1}\xi_{i_4}
\delta_{i_2 i_3} +\xi_{i_2}\xi_{i_3} \delta_{i_1 i_4}+ \xi_{i_2}\xi_{i_4}
\delta_{i_1 i_3} +\xi_{i_3}\xi_{i_4}\delta_{i_1 i_2} \big )+ \nonumber \\
&\,\delta_{i_1 i_2 i_3 i_4}
\end{flalign*}
The tensorial basis that spans the new generalized polynomials contains the basis that spans the Hermite polynomials but not vice-versa. The D-dimensional Hermite polynomials $\mathcal{P}_{i_1\cdots i_N}$ are symmetric tensors in the  indices $i_1\cdots i_N$ spanned over the basis formed by the tensors,
$$T_{i_1\cdots i_N}\equiv\xi_{i_1}\cdot \xi_{i_2}\cdots \xi_{i_P}\cdot \delta_{i_{P+1},i_{P+2}}\cdot\delta_{i_{P+3},i_{P+4}}\cdots\delta_{i_{N-1},i_{N}}.$$
This basis is not large enough to span the new generalized polynomials,
$\mathcal{P}_{i_1\cdots i_N}$, which demand a larger basis formed by the tensors
$$T_{i_1\cdots i_N}\equiv F\big( \xi\big) \xi_{i_1}\cdot \xi_{i_2}\cdots \xi_{i_P}\cdot \delta_{i_{P+1},i_{P+2}}\cdot \delta_{i_{P+3},i_{P+4}}\cdots\delta_{i_{N-1},i_{N}},$$
whose scalar functions $F\big( \xi \big)$ are polynomials in powers of the vector modulus, $1$, $\xi^2$, $\xi^4$, ...,$\xi^{2k}$.

\section{Expansion in polynomials orthonormal under a general weight}\label{general-weight}

In this section, we consider the series expansion in  polynomials orthonormal under a general weight  $\omega( \xi )$.
Although we are ultimately interested in the situation that the weight is similar to the EDF itself, $\omega( \xi ) \approx f^{(eq)}(\boldsymbol{\xi})$, because this makes convergence faster, we do no take this assumption here. For the special case that $\omega(\xi) = f^{eq}(\xi)$ some extra properties can be derived and this is done in the next section. The projections $\mathcal{A}_{i_1\,i_2 \cdots i_N}$  are obtained from the general orthonormal relation of Eq.(\ref{omeg-feq}),
\begin{flalign*}
\mathcal{A}_{i_1\,i_2 \cdots i_N}(\boldsymbol{u} )= \int d^D \boldsymbol{\xi^{\prime}} \, f^{(eq)}(\boldsymbol{\xi^{\prime}}-\boldsymbol{u}) \mathcal{P}_{i_1\,i_2\cdots i_N}( \boldsymbol{\xi^{\prime}}).\nonumber
\end{flalign*}
There is a  completeness relation for these generalized tensorial polynomials, which is obtained from the above expression and Eq.(\ref{feq-complete-exp-eq}):
\begin{flalign*}
& f^{(eq)}(\boldsymbol{\xi} -\boldsymbol{u})= \omega( \boldsymbol{\xi} )\cdot \sum_{N=0}^{\infty}\frac{1}{N!} \int d^D \boldsymbol{\xi^{\prime}} \, \nonumber\\
&f^{(eq)}(\boldsymbol{\xi^{\prime}}-\boldsymbol{u}) \mathcal{P}_{i_1\,i_2\cdots i_N}( \boldsymbol{\xi^{\prime}}) \mathcal{P}_{i_1\,i_2\cdots i_N}( \boldsymbol{\xi}),
\end{flalign*}
since $f(\boldsymbol{\xi}-\boldsymbol{u})= \int d^D \boldsymbol{\xi^{\prime}} \,\delta^{D}(\boldsymbol{\xi^{\prime}}-\boldsymbol{\xi})f(\boldsymbol{\xi^{\prime}}-\boldsymbol{u})$.
The completeness relation is given by,
\begin{flalign*}
 \omega( \boldsymbol{\xi} ) \sum_{N=0}^{\infty}\frac{1}{N!} \, \mathcal{P}_{i_1\,i_2\cdots i_N}( \boldsymbol{\xi^{\prime}}) \mathcal{P}_{i_1\,i_2\cdots i_N}( \boldsymbol{\xi}) = \delta^{D}(\boldsymbol{\xi^{\prime}}-\boldsymbol{\xi}),
\end{flalign*}

In terms of the relative (or peculiar) velocity $\boldsymbol{\eta}=\boldsymbol{\xi}-\boldsymbol{u}$ the quantities $\rho$, $\bar \theta$ and $g$ become,
\begin{flalign}
&\rho \equiv \int d^D \boldsymbol{\eta}f^{(eq)}(\boldsymbol{\eta}),\\
&\rho \bar\theta \delta_{i_1i_2} \equiv \int d^D \boldsymbol{\eta} f^{(eq)}(\boldsymbol{\eta}) \eta_{i_1}\eta_{i_2}, \\
& \rho {\bar\theta}^2 g \,\delta_{i_1i_2i_3i_4}\equiv \int d^D\boldsymbol{\eta} f^{(eq)}(\boldsymbol{\eta}) \eta_{i_1}\eta_{i_2}\eta_{i_3}\eta_{i_4}. \label{int-2nd-coeff-eq}
\end{flalign}
For the case of the FD-BE EDF these quantities can be expressed in terms of the integrals defined as
\begin{flalign}
g_\nu(z) = \frac{1}{\Gamma(\nu)}\int^\infty_0 dx\frac{x^{\nu -1}}{z^{-1}\exp(x) \pm 1},
\end{flalign}
which are functions of the fugacity $z$.
One obtains that,
\begin{flalign}
& \rho(z, \theta) = (2\pi\theta)^{\frac{D}{2}}g_{\frac{D}{2}}(z), \label{rho-fd-be-eq}\\
& \bar \theta(z, \theta) = \theta \frac{g_{\frac{D}{2}+1}(z)}{g_{\frac{D}{2}}(z)},\label{bar-theta-fd-be-eq}\\
& g(z) = \frac{g_{\frac{D}{2}}(z)g_{\frac{D}{2}+2}(z)}{\left( g_{\frac{D}{2}+1}(z)\right)^2}.\label{gz-fd-be-eq}
\end{flalign}
Recall that the EDF is solely a function of the modulus of the relative velocity, namely, $f^{eq}(\eta)$, $\eta = \vert \boldsymbol{\eta}\vert $. Below the first five projections of the EDF are calculated by noticing that by taking that  $d^D \boldsymbol{\xi}=d^D \boldsymbol{\eta}$ the limits of integration do not change  since the integrand vanishes exponentially at infinity.\\
\noindent $\bullet$ {\bf Zeroth order} -- Since $\mathcal{P}_0(\boldsymbol{\xi})=c_0$, one obtains that
\begin{flalign}
\mathcal{A}_0 = \int d^D \boldsymbol{\eta} f^{(eq)}(\boldsymbol{\eta}) \mathcal{P}_0(\boldsymbol{\xi})= c_0 \rho
\end{flalign}

\noindent $\bullet$ {\bf First order} -- Since
$\mathcal{P}_{i_1}(\boldsymbol{\xi}) = \mathcal{P}_{i_1}(\boldsymbol{\eta}) + c_1 u_{i_1}$,
there are two terms to consider for the projection,
\begin{flalign*}
\mathcal{A}_{i_1} = \int d^D \boldsymbol{\eta} f^{(eq)}(\boldsymbol{\eta}) \mathcal{P}_{i_1}(\boldsymbol{\xi}),
\end{flalign*}
The first term gives no contribution because the integration of an odd function vanishes. Hence only the second term contributes and gives that,
\begin{flalign}
\mathcal{A}_{i_1} = c_1 u_{i_1}\int d^D \boldsymbol{\eta} f^{(eq)}(\boldsymbol{\eta}) = \rho c_1 u_{i_1}.
\end{flalign}

\noindent $\bullet$ {\bf Second order} -- In the projection,
\begin{flalign*}
\mathcal{A}_{i_1i_2} = \int d^D \boldsymbol{\eta} f^{(eq)}(\boldsymbol{\eta}) \mathcal{P}_{i_1i_2}(\boldsymbol{\xi}),
\end{flalign*}
we introduce the expanded polynomial,
\begin{flalign*}
& \mathcal{P}_{i_1i_2}(\boldsymbol{\xi}) = \mathcal{P}_{i_1i_2}(\boldsymbol{\eta}) + \frac{c_2}{c_1} [u_{i_1} \mathcal{P}_{i_2}(\boldsymbol{\eta}) +  u_{i_2} \mathcal{P}_{i_1}(\boldsymbol{\eta})]\nonumber \\ & + 2 \frac{\bar c_2}{c_1} u_{i_3}\mathcal{P}_{i_3}(\boldsymbol{\eta})\delta_{i_1i_2}  + c_2 u_{i_1} u_{i_2} + \bar c_2 u^2 \delta_{i_1i_2}.
\end{flalign*}
The odd terms vanish and so,
\begin{flalign}
\mathcal{A}_{i_1i_2} &= \rho \left[ (c_2\bar \theta + c_2') \delta_{i_1i_2} + c_2 u_{i_1}u_{i_2} \right. \nonumber \\ &  \left. + \bar c_2(D \bar \theta  + \boldsymbol{u}^2) \delta_{i_1i_2}\right]
\end{flalign}

\noindent $\bullet$ {\bf Third order} -- Similarly, the projection,
\begin{flalign*}
\mathcal{A}_{i_1i_2i_3} = \int d^D \boldsymbol{\eta} f^{(eq)}(\boldsymbol{\eta}) \mathcal{P}_{i_1i_2i_3}(\boldsymbol{\xi})
\end{flalign*}
is calculated using the expanded polynomial.
\begin{flalign*}
&\mathcal{P}_{i_1i_2i_3}(\boldsymbol{\xi}) = \mathcal{P}_{i_1i_2i_3}(\boldsymbol{\eta}) + c_3 u_{i_1} \eta_{i_2}\eta_{i_3} + c_3 u_{i_3} \eta_{i_1} \eta_{i_2} \nonumber \\
&+ c_3 u_{i_1} u_{i_3} \eta_{i_2} + c_3 u_{i_2} \eta_{i_1}\eta_{i_3}+ c_3 u_{i_2} u_{i_3} \eta_{i_1} \nonumber \\
&+ c_3 u_{i_2} u_{i_3} u_{i_1}+ c_3 u_{i_2} u_{i_1} \eta_{i_3} + (\bar c_3 \eta^2 + c_3')(u_{i_1}\delta_{i_2i_3}
\nonumber \\
&+ u_{i_2} \delta_{i_1i_3}+ u_{i_3}\delta_{i_1i_2}) + \bar c_3 u^2(\eta_{i_1}\delta_{i_2i_3} + \eta_{i_2}\delta_{i_1i_3}
\nonumber \\
&+ \eta_{i_3}\delta_{i_1i_2} + u_{i_1}\delta_{i_2i_3}+ u_{i_2}\delta_{i_1i_3} + u_{i_3}\delta_{i_1i_2}) \nonumber \\
&+ 2 \bar c_3 u_{i_4}\eta_{i_4}(\eta_{i_1}\delta_{i_2i_3} + \eta_{i_2}\delta_{i_1i_3} + \eta_{i_3}\delta_{i_1i_2} + u_{i_1}\delta_{i_2i_3} \nonumber \\
&+ u_{i_2}\delta_{i_1i_3}+ u_{i_3} \delta_{i_1i_2})
\end{flalign*}
Using Eq.\eqref{int-2nd-coeff-eq} we get that,
\begin{flalign}
&\mathcal{A}_{i_1i_2i_3}= \rho \left[ \big(c_3\bar\theta + \bar c_3 \bar\theta (D+2) + c_3' + \bar c_3 u^2\big)\right. \nonumber \\
& \left.(u_{i_1}\delta_{i_2i_3} + u_{i_3}\delta_{i_1i_2}+ u_{i_2}\delta_{i_1i_3})+ c_3u_{i_1}u_{i_2}u_{i_3}\right].
\end{flalign}

\noindent $\bullet$ {\bf Fourth order} -- The last projection obtained here is,
\begin{flalign*}
\mathcal{A}_{i_1i_2i_3i_4} = \int d^D \boldsymbol{\eta} f^{(eq)}(\boldsymbol{\eta}) \mathcal{P}_{i_1i_2i_3i_4}(\boldsymbol{\xi})
\end{flalign*}
The expansion of the N=4 polynomial in the variable $\boldsymbol{\xi}=\boldsymbol{\eta}+\boldsymbol{u}$ renders a complex expression and for this reason we write below only its even terms, the only ones to contribute to the projection.
\begin{flalign*}
&\mathcal{P}_{i_1i_2i_3i_4}(\boldsymbol{\xi}) = c_4 (\eta_{i_1}\eta_{i_2}\eta_{i_3}\eta_{i_4}+ \eta_{i_1}\eta_{i_2}u_{i_3}u_{i_4} \nonumber \\
&+ \eta_{i_1}\eta_{i_3}u_{i_2}u_{i_4} + \eta_{i_1}\eta_{i_4}u_{i_2}u_{i_3}+\eta_{i_2}\eta_{i_3}u_{i_1}u_{i_4}\nonumber \\
&+ \eta_{i_2}\eta_{i_4}u_{i_1}u_{i_3}+ \eta_{i_3}\eta_{i_4}u_{i_1}u_{i_2}+u_{i_1}u_{i_2}u_{i_3}u_{i_4})\nonumber \\
& + (c_4'+ \bar c_4 \eta^2 + \bar c_4 u^2)[(\eta_{i_1}\eta_{i_2}+u_{i_1}u_{i_2})\delta_{i_3i_4}\nonumber \\
&+ (\eta_{i_1}\eta_{i_3}+ u_{i_1}u_{i_3})\delta_{i_2i_4}+ (\eta_{i_1}\eta_{i_4} + u_{i_1}u_{i_4})\delta_{i_2i_3} \nonumber \\
&+ (\eta_{i_3}\eta_{i_4}+ u_{i_3}u_{i_4})\delta_{i_1i_2} + (\eta_{i_2}\eta_{i_4} + u_{i_2}u_{i_4})\delta_{i_1i_3} \nonumber \\
&+ (\eta_{i_2}\eta_{i_3} + u_{i_2}u_{i_3})\delta_{i_1i_4}] + 2\bar c_4(\boldsymbol{\eta}\cdot \boldsymbol{u})[(\eta_{i_1}u_{i_2}\nonumber \\
&+ \eta_{i_2}u_{i_1})\delta_{i_3i_4} + (\eta_{i_1}u_{i_3}+\eta_{i_3}u_{i_1})\delta_{i_2i_4} + (\eta_{i_1}u_{i_4}\nonumber \\
&+ \eta_{i_4}u_{i_1})\delta_{i_2i_3} + (\eta_{i_3}u_{i_4} + \eta_{i_4}u_{i_3})\delta_{i_1i_2}+ (\eta_{i_2}u_{i_4}\nonumber \\
& + \eta_{i_4}u_{i_2})\delta_{i_1i_3}+ (\eta_{i_2}u_{i_3}+\eta_{i_3}u_{i_2})\delta_{i_1i_4}] + [d_4 \nonumber \\
&+ d_4'(\eta^2 + u^2) + \bar d_4 (\eta^4 + 2 \eta^2 u^2 + 4(\boldsymbol{\eta}\cdot \boldsymbol{u})^2)  \nonumber \\
&+ u^4] (\delta_{i_1i_2}\delta_{i_3i_4}+ \delta_{i_1i_3}\delta_{i_2i_4}+ \delta_{i_1i_4}\delta_{i_2i_3})\nonumber \\
& + \mbox{odd terms in $\eta$}.
\end{flalign*}
we have finally the fourth order projection:
\begin{flalign}
&\mathcal{A}_{i_1i_2i_3i_4} = \rho \Big\{  \delta_{i_1i_2i_3i_4}\big[ c_4{\bar\theta}^2 g + 2(c_4'+\bar c_4u^2)\bar\theta \nonumber \\
&+ 2\bar c_4 {\bar\theta}^2 g(D+2) + d_4 + d_4'(D\bar\theta+u^2) \nonumber \\
&+ \bar d_4[{\bar\theta}^2 g D(D+2) + 2u^2\bar\theta D + 4u^2\bar\theta + u^4]\big] \nonumber \\
&+ (\delta_{i_1i_2}u_{i_3}u_{i_4} + \delta_{i_1i_3}u_{i_2}u_{i_4} + \delta_{i_1i_4}u_{i_2}u_{i_3}+ \delta_{i_2i_3}u_{i_1}u_{i_4} \nonumber \\
&+ \delta_{i_2i_4}u_{i_1}u_{i_3} + \delta_{i_3i_4}u_{i_1}u_{i_2})(c_4\bar\theta + c_4' + \bar c_4 u^2 \nonumber \\
&+ \bar c_4 \bar \theta D + 4\bar c_4 \bar \theta) + c_4  u_{i_1}u_{i_2}u_{i_3}u_{i_4}\Big\}.
\end{flalign}
From this we obtain the series expansion of the EDF until fourth order.
\begin{flalign}
&f^{(eq)}(\boldsymbol{\xi} -\boldsymbol{u})= \omega( \boldsymbol{\xi} ) \big\{\mathcal{A}^0\mathcal{P}_0+
\mathcal{A}_{i_1}\mathcal{P}_{i_1}+\frac{1}{2}\mathcal{A}_{i_1i_2}\mathcal{P}_{i_1i_2} \nonumber \\
&+\frac{1}{6}
\mathcal{A}_{i_1i_2i_3}\mathcal{P}_{i_1i_2i_3}+
\frac{1}{24}\mathcal{A}_{i_1i_2i_3i_4}\mathcal{P}_{i_1i_2i_3i_4}\big\}. \label{feq-4order-01}
\end{flalign}
where
\begin{flalign*}
& \mathcal{A}_0\mathcal{P}_0 = \rho c_0^2, \nonumber \\
& \mathcal{A}_{i_1}\mathcal{P}_{i_1} = \rho c_1^2 (\boldsymbol{\xi}\cdot \boldsymbol{u}), \nonumber \\
&\mathcal{A}_{i_1i_2}\mathcal{P}_{i_1i_2} = \rho \{ c_2 (c_2\bar \theta + c_2')\xi^2 + c_2^2 (\boldsymbol{\xi}\cdot \boldsymbol{u})^2 \nonumber \\
&+ c_2\bar c_2 (D \bar \theta + u^2)\xi^2 + (\bar c_2 \xi^2 + c_2') [D (c_2\bar \theta\nonumber \\
& + c_2') + c_2 u^2 + \bar c_2 D (D\bar \theta + u^2)],\\
&\mathcal{A}_{i_1i_2i_3}\mathcal{P}_{i_1i_2i_3} = \rho \big\{ 3(c_3\bar \theta + \bar c_3 \bar \theta (D+2) + c_3' + \bar c_3 u^2 )\nonumber \\
&(\boldsymbol{\xi}\cdot \boldsymbol{u})[c_3 \xi^2 + (\bar c_3 \xi^2 + c_3')(D+2)] + c_3^2 (\boldsymbol{\xi} \cdot \boldsymbol{u})^3 \nonumber \\
&+ 3c_3 u^2 (\bar c_3 \xi^2 + c_3') (\boldsymbol{\xi}\cdot \boldsymbol{u})\big\},\, \mbox{and}\\
&\mathcal{A}_{i_1i_2i_3i_4}\mathcal{P}_{i_1i_2i_3i_4} = \rho \Big\{ \big[c_4 {\bar \theta}_2 g + 2(c_4' + \bar c_4 u^2) \bar \theta \nonumber \\
&+ 2\bar c_4 {\bar \theta }_2 g (D+2) + d_4 + d_4'(D\bar \theta + u^2) \nonumber \\
&+ \bar d_4 [ {\bar \theta}_2 g D(D+2) + 2u^2\bar \theta D + 4u^2\bar \theta + u^4]\big]\nonumber \\
& \cdot\big[ 3c_4\xi^4 +6(c_4' + \bar c_4 \xi^2)\xi^2(D+2) + 3(d_4+d_4'\xi^2 \nonumber \\
&+ \bar d_4\xi^4)D(D+2)\big] + [c_4\bar\theta + c_4' + \bar c_4u^2 + \bar c_4\bar \theta D \nonumber \\
&+ 4\bar c_4 \bar \theta] \big[  6c_4\xi^2(\boldsymbol{\xi} \cdot \boldsymbol{u})^2 + 6(c_4' + \bar c_4 \xi^2)[\xi^2u^2 \nonumber \\
&+ (\boldsymbol{\xi}\cdot \boldsymbol{u})^2(D+4)] + 6(d_4 + d_4'\xi^2 + \bar d_4 \xi^4)(D\nonumber \\
&+2)u^2\big] + c_4^2(\boldsymbol{\xi}\cdot \boldsymbol{u})^4 + 6c_4(c_4' + \bar c_4 \xi^2)(\boldsymbol{\xi}\cdot \boldsymbol{u})^2 u^2 \nonumber \\
& + 3(d_4 + d_4' \xi^2 + \bar d_4 \xi^4)u^4 c_4.
\end{flalign*}
The question concerning convergence boils down to know that the $u=0$ limit $f^{(eq)}(\xi)$ has a reliable expression given by $\omega(\xi)$ times a polynomial of fourth order in $\xi$. Obviously the minimum condition is that $\omega(\xi)$ be sufficiently close to $f^{(eq)}(\xi)$ otherwise it will not be possible.

The numerical simulations of  Sec. \ref{lbm-electrons-sec} are done with the EDF expanded to second order, and for this reason, we write it below.
\begin{flalign}\label{edf-second-order-exp-eq}
 &f^{(eq)}(\boldsymbol{\xi} -\boldsymbol{u}) = \rho \,\omega(\xi) \Big\{  c_0^2 + c_1^2(\boldsymbol{\xi}\cdot\boldsymbol{u}) + \frac{1}{2}c_2(c_2\bar \theta+ c_2')\xi^2 \nonumber\\ &
 + \frac{c_2^2}{2}(\boldsymbol{\xi}\cdot\boldsymbol{u})^2 + \frac{1}{2}c_2 \bar c_2 (D\bar \theta + u^2)\xi^2+\frac{1}{2}(\bar c_2\xi^2 + c_2')\nonumber \\ &
 \cdot [D(c_2\bar\theta+ c_2')+c_2 u^2+D\bar c_2 (D\bar \theta+u^2)] \Big\} .
\end{flalign}

\section{Expansion in polynomials orthonormal for $\omega(\xi)=f^{eq}(\xi)$}\label{special-weight}

In this section, we consider the series expansion in  polynomials orthonormal under a  weight  equal to the EDF itself,
\begin{flalign}
\omega(\xi) \equiv f^{(eq)}(\xi).
\label{omega-feq}
\end{flalign}
All the results of the previous section still holds, nevertheless the above choice for the weight brings special properties to the projections, such as,
\begin{flalign}
\sum_{N=0}^{\infty}\frac{1}{N!} \mathcal{A}_{i_1\,i_2 \cdots i_N}(0 ) \mathcal{P}_{i_1\,i_2\cdots i_N}( \boldsymbol{\xi})=1,
\label{sum-a0}
\end{flalign}
from where it follows that $\mathcal{A}_{i_1i_2\ldots i_N}(0)= 0$ for $N\geq 1$ since $\mathcal{A}_{0}(0)\mathcal{P}_{0}( \boldsymbol{\xi})=1$.
This has an important consequence for convergence, these projections are guaranteed to be small in case of a small macroscopic velocity $u \ll 1$. It holds that
$\mathcal{A}_{i_1i_2\ldots i_N}(\boldsymbol{u}) \approx u_{i_1} \delta_{i_2\ldots i_N} $ and $\mathcal{A}_{i_1i_2\ldots i_N}(\boldsymbol{u}) \approx u^2 \delta_{i_1 i_2\ldots i_N} + a. u_{i_1\,i_2} \delta_{i_3\ldots i_N}$, where $a$ is $a$ coefficient. This holds for $N$ odd and even, except in case of $\mathcal{A}_0$. Therefore  the choice of Eq.(\ref{omega-feq}) has important consequences, specially useful in case of the semiclassical statistics, given by the FD and BE EDFs~\cite{coelho16-2}.
We notice that the the macroscopic velocity $\boldsymbol{u}$ is a ratio normalized by a velocity appropriate to the bosons or fermions not necessarily equal to Mach's velocity.

The previously defined quantities $\rho$, $\bar \theta$ and $g$ become integrals defined in Eq.(\ref{i2n-cont}),
\begin{flalign}
&\rho \equiv \int d^D \boldsymbol{\eta}\omega(\boldsymbol{\eta})=I_0, \label{rho-i0}\\
&\rho \bar\theta \delta_{i_1i_2} \equiv \int d^D \boldsymbol{\eta} \omega(\boldsymbol{\eta}) \eta^{i_1}\eta^{i_2}=I_2 \delta_{i_1i_2}, \\
& \rho {\bar\theta}^2 g \,\delta_{i_1i_2i_3i_4}\equiv \int d^D\boldsymbol{\eta} \omega(\boldsymbol{\eta}) \eta^{i_1}\eta^{i_2}\eta^{i_3}\eta^{i_4}=I_4 \delta_{i_1i_2i_3i_4}. \label{int-2nd-coeff-eq2}
\end{flalign}
New and interesting expressions for the projections emerge by considering that $\mathcal{P}_0(\boldsymbol{\eta})/c_0=1$ and the fact that the weight function is equal to the EDF.
\begin{flalign}\label{acoeff2}
\mathcal{A}_{i_1\,i_2 \cdots i_N}(\boldsymbol{u} )= \frac{1}{c_0}\int d^D \boldsymbol{\eta} \, \omega(\boldsymbol{\eta}) \, \mathcal{P}_0(\boldsymbol{\eta}) \,\mathcal{P}_{i_1\,i_2\cdots i_N}( \boldsymbol{\eta}+\boldsymbol{u}),
\end{flalign}
Hence the determination of the projections is reduced to the expansion $\mathcal{P}_{i_1\,i_2\cdots i_N}( \boldsymbol{\eta}+\boldsymbol{u})$ as a sum over polynomials $\mathcal{P}_{i_1\,i_2\cdots i_M}( \boldsymbol{\eta})$ of equal or lower order ($M \le N$).
\begin{flalign*}
& \mathcal{P}_{i_1i_2\ldots i_N}(\boldsymbol{\eta}+\boldsymbol{u}) =  \nonumber \\  & \mathcal{U}_0(\boldsymbol{u})\mathcal{P}_{i_1i_2\ldots i_N}(\boldsymbol{\eta}) +\ldots+ \mathcal{U}_{i_1i_2\ldots i_N}(\boldsymbol{u})\mathcal{P}_{0}(\boldsymbol{\eta}).\quad \quad
\end{flalign*}
The coefficients $\mathcal{U}_{i_1i_2\ldots i_{N-M}}(\boldsymbol{u})$ that multiplies  the polynomial $\mathcal{P}_{i_1i_2\ldots i_M}(\boldsymbol{\eta})$ are tensors built from products of components $u_i$ and the Kronecker´s delta function $\delta_{i\,j}$ times
coefficients which are themselves polynomials in $u^2$ without the constant term with the exception of $\mathcal{U}_{0}$ which is a constant itself.
Indeed according to the above expression in the limit $\boldsymbol{u} \rightarrow 0$, it holds  that
$\mathcal{U}_0(0)=1$ while for the higher order tensors $\mathcal{U}_{i_1i_2\ldots i_M}(0)=0$. In summary the sought projections obtained from Eq.(\ref{acoeff2}) become,
\begin{flalign*}
\mathcal{A}_{i_1i_2\ldots i_N}(\boldsymbol{u})=\frac{1}{c_0} \mathcal{U}_{i_1i_2\ldots i_N}(\boldsymbol{u}).
\end{flalign*}
Then it follows that $\mathcal{A}_{i_1 i_2\ldots i_N}(\boldsymbol{u}=0)= 0$ for $N\geq 1$, as previously stated. The obtainment of  the projections in case the weight is the EDF itself  is reduced to calculate the polynomial expansion, and we do it explicitly to order N=4.\\
\noindent $\bullet$ {\bf  Zeroth order} -- The  expansion in case of N=0 is $\mathcal{P}_{0}(\boldsymbol{\eta} +\boldsymbol{u})=\mathcal{U}_0\mathcal{P}_0(\boldsymbol{\eta})$, hence $\mathcal{U}_0=1$.\\
\noindent $\bullet$ {\bf  First order} -- The expansion in case of N=1 is
$\mathcal{P}_{i_1} (\boldsymbol{\eta} + \boldsymbol{u}) = c_1(\eta_{i_1}+u_{i_1})=
\mathcal{U}_0(\boldsymbol{u})\mathcal{P}_{i_1}(\boldsymbol{\eta})+ \mathcal{U}_{i_1}(\boldsymbol{u})\mathcal{P}_{0}(\boldsymbol{\eta})$.
Thus $\mathcal{U}_0(\boldsymbol{u})=1$ and
\begin{flalign*}
\mathcal{U}_{i_1}(\boldsymbol{u})=\frac{c_1}{c_0} u_{i_1}.\nonumber
\end{flalign*}
\noindent $\bullet$ {\bf  Second order} -- Expanding the N=2 polynomial,
$\mathcal{P}_{i_1i_2}(\boldsymbol{\eta}+\boldsymbol{u}) = c_2 (\eta_{i_1} + u_{i_1})(\eta_{i_2}+u_{i_2}) \nonumber + [\bar c_2 (\boldsymbol{\eta}+ \boldsymbol{u})^2 + c^{\prime}_2] \delta_{i_1i_2}$, gives that,
\begin{flalign*}
&\mathcal{P}_{i_1i_2}(\boldsymbol{\eta}+\boldsymbol{u})= \mathcal{P}_{i_1i_2}(\boldsymbol{\eta}) + \nonumber \\ &\frac{c_2}{c_1}[u_{i_1}\mathcal{P}_{i_2}(\boldsymbol{\eta}) + u_{i_2}\mathcal{P}_{i_1}(\boldsymbol{\eta})] + \frac{2\bar c_2}{c_1} u_{i_3}\mathcal{P}_{i_3}(\boldsymbol{\eta}) \delta_{i_1i_2}\nonumber \\
&+ \frac{1}{c_0}\mathcal{P}_0(\boldsymbol{\eta})\left[c_2 u_{i_1}u_{i_2} +\bar c_2 \boldsymbol{u}^2 \delta_{i_1i_2} \right].
\end{flalign*}
Therefore one obtains that,
\begin{flalign*}
\mathcal{U}_{i_1 i_2}(\boldsymbol{u})= \frac{1}{c_0}(c_2 u_{i_1}u_{i_2} +\bar c_2 \boldsymbol{u}^2 \delta_{i_1i_2}). \nonumber
\end{flalign*}
\noindent $\bullet$ {\bf Third order} --Similarly the expansion of the N=3 polynomial, $\mathcal{P}_{i_1i_2i_3}(\boldsymbol{\eta} + \boldsymbol{u})$, contains the N=0 polynomial plus  higher order ones that are omitted for simplicity.
\begin{flalign*}
&\mathcal{P}_{i_1i_2i_3}(\boldsymbol{\eta}+\boldsymbol{u}) = \frac{\mathcal{P}_0(\boldsymbol{\eta})}{c_0} \Big \{ \Big[ \frac{c_3 \bar c_2 D c^{\prime}_2}{c_2(c_2 + D\bar c_2)} -  \frac{c_3 c^{\prime}_2}{c_2}- \nonumber \\
&\frac{\bar c_3 D c^{\prime}_2}{(c_2+ D\bar c_2)}+ c^{\prime}_3+\bar c_3\boldsymbol{u}^2 + \frac{2\bar c_3 \bar c_2 D c^{\prime}_2}{c_2(c_2+ D\bar c_2)} - \frac{2\bar c_3 c^{\prime}_2}{c_2} \Big ] \nonumber \\
& u_{i_4}\delta_{i_1i_2i_3i_4} + c_3u_{i_1}u_{i_2}u_{i_3}  \Big \} + \mathcal{O}(\boldsymbol{\eta})
\end{flalign*}
The above equation can be simplified using the expressions of the coefficients.
\begin{flalign*}
&\mathcal{P}_{i_1i_2i_3}(\boldsymbol{\eta}+\boldsymbol{u}) = \mathcal{P}_0(\boldsymbol{\eta}) \mathcal{U}_{i_1i_2i_3}(\boldsymbol{u})+ \mathcal{O}(\boldsymbol{\eta}), \nonumber \\
& \mathcal{U}_{i_1i_2i_3}(\boldsymbol{u})= \frac{1}{c_0}\Big\{\Big[\frac{I_2}{I_0}[c_3+\bar c_3(D+2)] + \bar c_3\boldsymbol{u}^2 + c^{\prime}_3\Big ]\nonumber \\
&u_{i_4}\delta_{i_1i_2i_3i_4} +  c_3u_{i_1}u_{i_2}u_{i_3}  \Big \}
\end{flalign*}
\noindent $\bullet$ {\bf Fourth order} -- The N=4 polynomial $\mathcal{P}_{i_1i_2i_3i_4}(\boldsymbol{\eta}+\boldsymbol{u})$ can be expanded in powers of $\boldsymbol{\eta}$ and such powers rearranged as a sum over the polynomials
$\mathcal{P}_{i_1 \cdots i_M}(\boldsymbol{\eta})$, M=0 to 4, must be . Nevertheless we only seek the N=0 term and some considerations can be applied to simplify this task. For instance, the odd terms ($\eta_{i_1}$, $\eta_{i_1}\boldsymbol{\eta}^2$, $\eta_{i_1}\eta_{i_2}\eta_{i_3}$) do not contribute to the calculation of $\mathcal{A}_{i_1i_2i_3i_4}$ and one can take that $\boldsymbol{\eta}^2 = DI_2/I_0 + \mathcal{O}(\boldsymbol{\eta})$. After some algebra, we have that:
\begin{flalign*}
&\mathcal{P}_{i_1i_2i_3i_4}(\boldsymbol{\eta}+\boldsymbol{u}) = \mathcal{P}_0(\boldsymbol{\eta})
 \mathcal{U}_{i_1i_2i_3i_4}(\boldsymbol{u})+ \mathcal{O}(\boldsymbol{\eta}), \nonumber \\
& \mathcal{U}_{i_1i_2i_3i_4}(\boldsymbol{u})= \frac{1}{c_0}\Big \{c_4 u_{i_1}u_{i_2}u_{i_3}u_{i_4} +
\Big[ \frac{I_2}{I_0}[c_4\nonumber \\
&+\bar c_4(D+4)] + c_4' + \bar c_4 \boldsymbol{u}^2\Big] (\delta_{i_1i_2}u_{i_3}u_{i_4} + \nonumber \\
&\delta_{i_1i_3}u_{i_2}u_{i_4}+ \delta_{i_1i_4}u_{i_2}u_{i_3}+ \delta_{i_2i_3}u_{i_1}u_{i_4} + \delta_{i_2i_4}u_{i_1}u_{i_3}+ \nonumber \\
& \delta_{i_3i_4}u_{i_1}u_{i_2}) +\Big [ 2\bar c_4 \frac{I_2}{I_0}\boldsymbol{u}^2 + d^{\prime}_4 \boldsymbol{u}^2  + 2D\frac{I_2}{I_0} \bar d_4 \boldsymbol{u}^2 + \nonumber \\
&4\frac{I_2}{I_0}\bar d_4 \boldsymbol{u}^2 + \boldsymbol{u}^4 \bar d_4\Big ]\delta_{i_1i_2i_3i_4} \Big \}.
\end{flalign*}
Further simplification gives that,
\begin{flalign*}
&\mathcal{U}_{i_1i_2i_3i_4}(\boldsymbol{u}) = \frac{1}{c_0} \Big\{c_4 u_{i_1}u_{i_2}u_{i_3}u_{i_4} + \Big[\frac{I_2}{I_0}[c_4+\bar c_4(D+4)] \nonumber \\
&+ c_4' + \bar c_4 \boldsymbol{u}^2\Big ](\delta_{i_1i_2}u_{i_3}u_{i_4} + \delta_{i_1i_3}u_{i_2}u_{i_4}+ \delta_{i_1i_4}u_{i_2}u_{i_3}+ \nonumber \\
&\delta_{i_2i_3}u_{i_1}u_{i_4}+\delta_{i_2i_4}u_{i_1}u_{i_3} +\delta_{i_3i_4}u_{i_1}u_{i_2}) +\Big[ 2\bar c_4 \frac{I_2}{I_0}\boldsymbol{u}^2 \nonumber \\
& +  d^{\prime}_4 \boldsymbol{u}^2 +2D\frac{I_2}{I_0} \bar d_4 \boldsymbol{u}^2 + 4\frac{I_2}{I_0}\bar d_4 \boldsymbol{u}^2 + \boldsymbol{u}^4 \bar d_4\Big ] \delta_{i_1i_2i_3i_4}\Big\}
\end{flalign*}
We summarize the projections below, obtained after some additional algebraic manipulation. Notice that they are functions of the integrals $I_{2N}$ previously defined.
\begin{flalign*}
&\mathcal{A}_0(\boldsymbol{u}) = I_0 c_0,\\
&\mathcal{A}_{i_1}(\boldsymbol{u})=I_0 c_1\,u_{i_1},\\
&\mathcal{A}_{i_1 i_2}(\boldsymbol{u})= I_0 \big ( c_2 u_{i_1} u_{i_2} + {\bar c}_2 \boldsymbol{u}^2 \,\delta_{i_1 i_2} \big ),\\
&\mathcal{A}_{i_1 i_2 i_3}(\boldsymbol{u})=I_0\big \{ c_3 \,u_{i_1} u_{i_2}u_{i_3}  +\big [ {c^\prime}_3 \big(1-J_2 \big) \nonumber \\
&+{\bar c}_3\boldsymbol{u}^2\big]\big (u_{i_1} \delta_{i_2 i_3} + u_{i_2}\delta_{i_1 i_3} +u_{i_3} \delta_{i_1 i_2}\big ) \big \}, \quad \mbox{and},  \\
&\mathcal{A}_{i_1 i_2 i_3 i_4}(\boldsymbol{u})=I_0\Big\{{c}_4 \,u_{i_1} u_{i_2}u_{i_3} u_{i_4}+ \big [\big( 1-J_2J_4\big){c^\prime}_4 \nonumber \\
&+{\bar c}_4 \boldsymbol{u}^2\big]\big ) \big (u_{i_1}u_{i_2}\delta_{i_3 i_4} + u_{i_1}u_{i_3}
\delta_{i_2 i_4}+u_{i_1}u_{i_4}\delta_{i_2 i_3}+ \nonumber \\
& u_{i_2}u_{i_3} \delta_{i_1 i_4}+ u_{i_2}u_{i_4} \delta_{i_2 i_3}+u_{i_3}u_{i_4}\delta_{i_1 i_2}\big ) + \big[ \big(2\frac{I_2}{I_0}\big( \bar c_4\nonumber\\
&+(D+2)\bar d_4\big) +d^\prime_4\big ) \boldsymbol{u}^2+\bar d_4 \boldsymbol{u}^4  \big ]\,\delta_{i_1 i_2 i_3 i_4}\Big \}.
\end{flalign*}
Using the definitions of the coefficients, one obtains that,
\begin{flalign*}
&\mathcal{A}_0 \mathcal{P}_0 = 1\\
&\mathcal{A}_{i_1} \mathcal{P}_{i_1} = \frac{I_0}{I_2} (\boldsymbol{\xi}\cdot\boldsymbol{u})\\
&\mathcal{A}_{i_1i_2} \mathcal{P}_{i_1i_2} = I_0\Big[\frac{1}{I_4}(\boldsymbol{\xi}\cdot\boldsymbol{u})^2 - \frac{(\Delta_2^2-1)}{I_4D}\boldsymbol{u}^2\boldsymbol{\xi}^2 - \frac{I_2}{I_0I_4}\Delta_2^2\boldsymbol{u}^2 \Big]\\
&\mathcal{A}_{i_1i_2i_3} \mathcal{P}_{i_1i_2i_3} = I_0 (\boldsymbol{\xi}\cdot\boldsymbol{u})\Big [3(1-J_2)\frac{J_4}{I_2}(D+2)\Delta_4^2 \\ & - 3\frac{J_4}{I_4}\Delta_4^2 \boldsymbol{u}^2- 3 (1-J_2)\frac{J_4}{I_4}\Delta_4^2 \boldsymbol{\xi}^2  + 3\frac{\Delta_4^2-1}{I_6(D+2)}\boldsymbol{\xi}^2\boldsymbol{u}^2 \\ & +  \frac{1}{I_6}(\boldsymbol{\xi}\cdot\boldsymbol{u})^2\Big]  \\
&\mathcal{A}_{i_1i_2i_3i_4} \mathcal{P}_{i_1i_2i_3i_4} = I_0 \Big\{ c_4^2 (\boldsymbol{\xi}\cdot\boldsymbol{u})^4 + 6c_4(c_4' + \bar c_4 \xi^2) u^2(\boldsymbol{\xi}\cdot \boldsymbol{u})^2 \\ &
+ 3c_4 u^4 (d_4 + d_4'\xi^2 + \bar d_4 \xi^4) + 6\big[ \frac{I_2}{I_0}(c_4+ \bar c_4(D+4)) \\ &
+ c_4' + \bar c_4 u^2 \big]\big[ c_4 \xi^2 (\boldsymbol{\xi}\cdot \boldsymbol{u})^2 + (c^{\prime}_4 +\bar c_4 \xi^2) [\xi^2 u^2 \\ &
+ (\boldsymbol{\xi}\cdot \boldsymbol{u})^2(D+4)]  + (d_4 +d_4'\xi^2 + \bar d_4 \xi^4)u^2(D+2)\big] \\ &
+ 3\big[ u^2 \frac{I_2}{I_0} 2(\bar c_4 + D \bar d_4 + 2\bar d_4)+ d_4' u^2 + \bar d_4 u^4 \big] [c_4\xi^4   \\ &
+  2(c_4' + \bar c_4 \xi^2) \xi^2 (D+2) + (d_4 + d_4' \xi^2 \\ &
+ \bar d_4 \xi^4)D (D+2)] \Big\}
\end{flalign*}

 Notice that the macroscopic velocity controls the smallness of the coefficients $\mathcal{A}_{i_1\,i_2 \cdots i_N}$, which to the lowest order are linear and quadratic in $\boldsymbol{u}$ for the odd and even ($N>0$) coefficients, respectively.

\subsection{Direct derivation of the equilibrium distribution function to order N=2 }\label{edfn2}
The EDF expanded to N=2 in case that $\omega(\xi) \equiv f^{(eq)}(\xi)$ is readily obtained from the sum of the first three above coefficients.
Here we derive this N=2 EDF assuming that it is a sum over all possible terms until the second power in the macroscopic velocity, namely,
$\boldsymbol{\xi}\cdot \boldsymbol{u}$, $(\boldsymbol{\xi}\cdot \boldsymbol{u})^2$, $\boldsymbol{u}^2$,  $\boldsymbol{\xi}^2 \boldsymbol{u}^2$.
\begin{flalign*}
f^{(eq)}=& \omega(\xi)\Big[f_0+f_1\,\boldsymbol{\xi}\cdot \boldsymbol{u}+\frac{f_2}{2}\,(\boldsymbol{\xi}\cdot \boldsymbol{u})^2 \\ &+\frac{f_3}{2}\, \boldsymbol{u}^2+\frac{f_4}{2}\,\boldsymbol{\xi}^2 \boldsymbol{u}^2\Big ]
\end{flalign*}
We find the coefficients $f_0$, $f_1$, $f_2$, $f_3$ and $f_4$ in the EDF given below, without invoking the orthonormal polynomials and just derive them from the given physical parameters, namely, the density, the macroscopic velocity and the temperature (Eqs.\eqref{rho-def-eq}, \eqref{vel-def-eq} and \eqref{theta-bar-def-eq}). Notice that only  integrals up to order $\xi^4$ are used in this derivation. Therefore the following relations are employed in the determination of the coefficients.
\begin{flalign*}
&\int d^D \boldsymbol{\xi} \omega (\xi)= I_0, \\
&\int d^D \boldsymbol{\xi} \omega (\xi) \, \xi_{ i_1}= 0,\\
&\int d^D \boldsymbol{\xi} \omega (\xi) \, \xi_{ i_1}\, \xi_{ i_2}= I_2 \, \delta_{i_1 i_2},\\
&\int d^D \boldsymbol{\xi} \omega (\xi) \, \xi_{ i_1} \,\xi_{ i_2} \, \xi_{ i_3}= 0,\\
&\int d^D \boldsymbol{\xi} \omega (\xi) \, \xi_{ i_1} \, \xi_{ i_2} \, \xi_{ i_3} \, \xi_{ i_4}= I_4 \, \delta_{i_1 i_2 i_3 i_4}.
\end{flalign*}
We start by imposing that Eq.(\ref{rho-def-eq}) holds to find that $\rho=I_0 f_0 + (f_2I_2+f_3I_0+f_4DI_2) \boldsymbol{u}^2/2$. Similarly from Eq.(\ref{vel-def-eq}) it follows that $\rho \boldsymbol{u}= f_1 I_2 \boldsymbol{u}$. Finally from Eq.(\ref{theta-bar-def-eq}) one obtains that $\rho ( u_{i_1} u_{i_2} + \bar \theta \delta_{i_1 i_2} )=  f_0I_2 \delta_{i_1 i_2}+f_2I_4 u_{i_1} u_{i_2} + (\boldsymbol{u}^2/2)\delta_{i_1 i_2}[f_2I_4+f_3I_2+(D+2)I_4 f_4]$.
Therefore
\begin{flalign*}
& \rho = I_0 f_0, \:\:\:
 f_2 I_2+f_3 I_0+f_4 D I_2=0,\\
& \rho = f_1 I_2, \:\:\:
 \rho \bar \theta = f_0 I_2, \:\:\:
 \rho = f_2 I_4,\\
& f_2 I_4 + f_3 I_2 + f_4 (D+2) I_4 =0.
\end{flalign*}
Solving these equations one obtains that $f_0=\rho/I_0$, $f_1=\rho/I_2$, $f_2=\rho/I_4$ and $\bar \theta = I_2/I_0$. Then one is left with the a system of equations to solve for the remaining two coefficients whose solution is $f_3=-(\rho I_2/I_4I_0) \Delta_2^2$ and $f_4 = (\rho/I_4)(\Delta_2^2-1)/D$ where $\Delta_2^2 = 2/[(D+2)-J_2D]$, $J_2=I_2^2/I_4 I_0$, as previously defined.
Finally one obtains that,
\begin{flalign}\label{feq-reduce-general-method-without-pol-eq}
&f^{(eq)}=\frac{\rho}{I_0}\omega(\xi) \Big \{1+ \frac{I_0}{I_2}\boldsymbol{\xi}\cdot \boldsymbol{u} +
 \frac{I_0}{I_4}\frac{1}{2}\,(\boldsymbol{\xi}\cdot \boldsymbol{u})^2 \nonumber \\
& + \frac{1}{2}\boldsymbol{u}^2\big[\frac{I_0}{I_4}\big(\frac{\Delta_2^2-1}{D}\big)
\boldsymbol{\xi}^2-\frac{I_2}{I_4}\Delta_2^2\big] \Big\}
\end{flalign}
This is equivalent to Eq.\eqref{edf-second-order-exp-eq} by substitution of the polynomial coefficients for the case $\omega(\xi)=f^{eq}(\xi)$ and taking that $\rho=I_0$, as given by Eq.\eqref{rho-i0}.

\section{Macroscopic equations}\label{hydrodynamic-sec}

In this section, we show the macroscopic equations followed by the semiclassical fluids, i.e., continuity, momentum conservation and energy conservation equations, and generalize their derivation, done in Ref.~\cite{coelho14}, for a generic EDF. We also discuss the minimum order that the EDF should be expanded in order to fully recover each macroscopic equation though the Chapman-Enskog expansion, developed below.

\subsection{General equations}\label{general-mac-eq-sec}

Here we list the general moments of the EDF needed to calculate the macroscopic equations. To recover the mass conservation (continuity equation):
\begin{flalign}
\frac{\partial \rho}{\partial t} + \frac{\partial}{\partial x_{i_1}}(\rho u_{i_1}) = 0
\label{continuity-eq}
\end{flalign}
 the zeroth and first order moments of the EDF are needed:
\begin{flalign}
\rho = \int d^D \xi f^{eq},  \:\:\:\:\rho u_{i_1} =\int d^D \xi \xi_{i_1} f^{eq}
\label{rho-int-eq}
\end{flalign}
To obtain the momentum equation, one needs to calculate the following second and third order moments:
\begin{flalign}
&\pi_{i_1i_2}=\int d^D \xi f^{eq} \xi_{i_1}\xi_{i_2},
\label{pi2-eq} \\
&
\pi_{i_1i_2i_3}=\int d^D \xi f^{eq} \xi_{i_1}\xi_{i_2} \xi_{i_3},
\label{pi3-eq}
\end{flalign}
which are subsequently introduced into the generic momentum equation:
\begin{flalign}
&\frac{\partial}{\partial t}(\rho u_{i_1}) + \frac{\partial}{\partial x_{i_2}}\pi_{i_1i_2} - \left(\tau - \frac{\Delta t}{2}\right) \frac{\partial}{\partial x_{i_2}}\frac{\partial}{\partial x_{i_3}}\pi_{i_1i_2i_3} \nonumber \\ &
-\left( \tau - \frac{\Delta t}{2}\right) \frac{\partial}{\partial x_{i_2}}\frac{\partial}{\partial t}\pi_{i_1i_2} = 0
\label{master-momentum-eq}
\end{flalign}

And for energy conservation equation, the moments needed are the second, third and fourth order ones:
\begin{flalign}
&\phi = \frac{1}{2}d^D\xi f^{eq} \xi^2 ,
\label{phi0-eq} \\
&
\phi_{i_1} = \frac{1}{2}d^D\xi f^{eq} \xi^2 \xi_{i_1} ,
\label{phi1-eq}\\
&
\phi_{i_1i_2} = \frac{1}{2}d^D\xi f^{eq} \xi^2 \xi_{i_1} \xi_{i_2}.
\label{phi2-eq}
\end{flalign}
which are introduced into the generic energy equation:
\begin{flalign}
&\frac{\partial}{\partial t}\phi + \frac{\partial}{\partial x_{i_1}}\phi_{i_1} - \left(  \tau - \frac{\Delta t}{2} \right) \frac{\partial }{\partial x_{i_1}}\frac{\partial}{\partial x_{i_2}}\phi_{i_1i_2} \nonumber \\  &- \left( \tau - \frac{\Delta t}{2}  \right)  \frac{\partial}{\partial x_{i_1}}\frac{\partial}{\partial t}\phi_{i_1} = 0  .
\label{master-energy-eq}
\end{flalign}
The Eqs. \eqref{continuity-eq}, \eqref{master-momentum-eq} and \eqref{master-energy-eq} give the macroscopic equations after calculating the moments above for a specific EDF and after some algebraic manipulations~\cite{coelho14, coelho2014lattice}.

\subsection{Macroscopic equations obtained with original function}\label{mac-eq-original-sec}

In Ref.~\cite{coelho14} the moments are calculated using the FD and BE EDFs expanded in Hermite polynomials up to fourth order. We generalize here this derivation for a generic non-expanded EDF. Using the definitions of $\rho$, $\boldsymbol{u}$, $\bar \theta$ and $g$ (Eqs. \eqref{rho-def-eq}, \eqref{vel-def-eq}, \eqref{theta-bar-def-eq} and \eqref{gz-def-eq} respectively), the moments can be straightforwardly calculated giving that:
\begin{flalign}
&\pi_{i_1i_2}=\rho \left[\bar \theta \delta_{i_1i_2} + u_{i_1}u_{i_2}\right], \nonumber\\
&\pi_{i_1i_2i_3}=\rho\Big[\bar\theta (u_{i_1}\delta_{i_2i_3}+u_{i_2}\delta_{i_1i_3}+u_{i_3}\delta_{i_1i_2})+u_{i_1}u_{i_2}u_{i_3}\Big],\nonumber
 \\
&\phi=\rho\left( \bar\theta \frac{D}{2} + \boldsymbol{u}^2 \right),\label{moments-non-expanded-edf-eq}\\
&
\phi_{i_1}=\frac{\rho}{2} u_{i_1} \left[\bar\theta (D+2) + \boldsymbol{u}^2\right],
\nonumber \\
&
\phi_{i_1i_2}=\frac{\rho}{2}\Big[(D+2) \delta_{i_1i_2} {\bar\theta}^2g + \bar\theta (D+4) u_{i_1}u_{i_2} \nonumber \\
&+ \bar\theta \boldsymbol{u}^2 \delta_{i_1i_2} + \boldsymbol{u}^2 u_{i_1} u_{i_2}\Big],\nonumber
\end{flalign}
which are the same ones found using the truncated fourth order expansion of EDF in Hermite polynomials~\cite{coelho14}. As we will show in the next section, terms from expansion orders higher than the monomial order in the integrand (that is,  $\xi_{i_1}\ldots \xi_{i_N}$, where $N$ is the monomial order) do not contribute to the moment because of the orthogonality of the polynomials. In addition, the results above would be the same if the EDF were expanded up to fourth order in any set of orthogonal polynomials.

Therefore, when the moments given in Eq.\eqref{moments-non-expanded-edf-eq} are substituted into  Eqs.\eqref{master-momentum-eq} and \eqref{master-energy-eq}, they give the same macroscopic equations obtained in Ref.~\cite{coelho14}. The momentum conservation equation or \textit{semiclassical Navier-Stokes} reads:
\begin{flalign}\label{semiclassical-NS-eq}
\frac{\partial}{\partial t}(\rho u_{i_1}) + \frac{\partial}{\partial x_{i_2}}[\rho(\bar\theta\delta_{i_1i_2} + u_{i_1}u_{i_2})] - \frac{\partial \bar\sigma_{i_1i_2}}{\partial x_{i_2}}=0.
\end{flalign}
And the energy conservation equation is given by:
\begin{flalign}\label{energy-cons-eq}
&\frac{\partial}{\partial t}\left[  \frac{\rho}{2}(u^2+D\bar\theta) \right] + \frac{\partial}{\partial x_{i_1}}\left[ \frac{\rho}{2}\left(u^2+\bar\theta(D+2) \right)u_{i_1} \right.\nonumber\\ & \left.+\tilde Q_{i_1} - u_{i_2}\bar \sigma_{i_1i_2}  \right] = 0,
\end{flalign}
where
\begin{flalign*}
\bar\sigma_{i_1i_2} = \bar \eta \left( \frac{\partial u_{i_1}}{\partial x_{i_2}} + \frac{\partial u_{i_2}}{\partial x_{i_1}}-\frac{2}{D}\delta_{i_1i_2}\frac{\partial u_{i_3}}{\partial x_{i_3}}\right)
\end{flalign*}
is the viscosity stress tensor and
\begin{flalign}
\tilde Q_{i_1} = - \bar \kappa\frac{\partial \bar \theta}{\partial x_{i_1}} - \frac{\partial}{\partial x_{i_1}} \left[\bar \kappa \bar\theta \left( g-1 \right)  \right]
\label{pseudo-heat-flux-eq}
\end{flalign}
is the heat flux vector (note that the second term disappears for the classical case, since $g=1$). Here the shear viscosity stands for $\bar \eta  = \rho\bar\theta \left(\tau - \frac{\Delta t}{2} \right)$ and the thermal conductivity for $\bar \kappa  = \frac{D+2}{2}\rho\bar\theta \left(\tau - \frac{\Delta t}{2} \right)$. When the EDF is the BE or the FD distribution, the Eq.\eqref{pseudo-heat-flux-eq} can be written as a function of the physical temperature and chemical potential gradients. In this case, the quantities $\bar\theta$ and $g(z)$ are given by Eqs.\eqref{bar-theta-fd-be-eq} and \eqref{gz-fd-be-eq} and Eq.\eqref{pseudo-heat-flux-eq} becomes:
\begin{flalign*}
\tilde Q_{i_1} = - \kappa_\theta \frac{\partial \theta}{\partial x_{i_1}} - \kappa _\mu \frac{\partial \mu}{\partial x_{i_1}},
\end{flalign*}
giving the following transport coefficient:
\begin{flalign*}
&\kappa_\theta = \kappa \left[\left(  \frac{D}{2} +2\right) \frac{g_{\frac{D}{2}+2}(z)}{g_{\frac{D}{2}}(z)} - \left(\frac{D}{2} +1  \right)\left( \frac{g_{\frac{D}{2}+1}(z)}{g_{\frac{D}{2}}(z)}  \right)^2 \right] \\ &
\kappa _\mu = 0,
\end{flalign*}
as obtained in Ref.~\cite{coelho2014lattice}.

\subsection{Macroscopic equations obtained with expanded EDF}

Here we analyze the minimum number of terms in the series expansion of the EDF that must be kept to obtain the macroscopic equations.
This amounts to determine $K$ in the series expansion defined in Sec.~\ref{edf-expansion-sec}.
According to Sec.~\ref{general-mac-eq-sec} the moments of the EDF, expressed by Eqs.\eqref{pi2-eq}, \eqref{pi3-eq}, \eqref{phi0-eq}, \eqref{phi1-eq} and \eqref{phi2-eq} must be obtained. They are integrals of the EDF multiplied by monomials of $\xi$.
For this reason we carry the following general discussion about a monomial tensor of order $N$, formed by the product of $N$ velocity components. It can be written as a sum over the generalized polynomials that  ranges from order zero to order $N$, that is,
\begin{flalign*}
\xi_{i_1}\xi_{i_2}\ldots\xi_{i_N} = C_0\mathcal{P}_{i_1\ldots i_N}+ \ldots+C_{i_1\ldots i_N}\mathcal{P}_0,
\end{flalign*}
where $C_{i_1\ldots i_N}$ are tensors constructed from the Kronecker's delta function and polynomials coefficients.  Thus, a moment of order $N$ of the expanded EDF gives:
\begin{flalign*}
&\pi_{i_1\ldots i_N} = \int d^D \xi f^{eq} \xi_{i_1}\ldots \xi_{i_N} \\ &
= \int d^D \xi \Big[ \rho\omega( \boldsymbol{\xi} ) \sum_{M=0}^{K}\frac{1}{M!} \mathcal{A}_{i_1 \cdots i_M}(\boldsymbol{u} ) \mathcal{P}_{i_1\cdots i_M}( \boldsymbol{\xi}) \Big] \\& \cdot \big[ C_0\mathcal{P}_{i_1\ldots i_N}+ \ldots+C_{i_1\ldots i_N}\mathcal{P}_0\big] .
\end{flalign*}
The importance of the orthogonality of the polynomials, Eq.~\eqref{omeg-feq}, can be appreciated at this point. The integrals of the terms in the series expansion which are of order $N+1$ and above simply vanish. This means that, a moment of order $N$ can be equally obtained either using the full non expanded EDF or  the expanded EDF  to order $N$.

For instance, to recover the continuity equation, Eq.\eqref{continuity-eq}, we need the zero and first order moments (Eq.\eqref{rho-int-eq}). Using that $1 = \mathcal{P}_0/c_0$ and $\xi_{i_1}=\mathcal{P}_{i_1}/c_1$, they become:
\begin{flalign*}
&\int d^D \xi \frac{\mathcal{P}_0}{c_0}f^{eq} = \frac{\mathcal{A}_0}{c_0} = \rho, \\ &
\int d^D \xi \frac{\mathcal{P}_{i_1}}{c_1} f^{eq} = \frac{\mathcal{A}_{i_1}}{c_1} = \rho u_{i_1}.
\end{flalign*}
Therefore, the \textit{first order} ($K=1$) expansion of the EDF is enough to recover the continuity equation and terms higher than this ($K \ge 2$) do not contribute.

For the momentum conservation equation, Eq.\eqref{semiclassical-NS-eq}, we need the second and third moments ($K=3$) given in Eqs.\eqref{pi2-eq} and \eqref{pi3-eq}. The monomials that are being integrated with the EDF can be written in terms of the polynomials as following:
\begin{flalign*}
&\xi_{i_1}\xi_{i_2} = \frac{1}{c_2}\mathcal{P}_{i_1i_2}-\frac{\bar c_2 \delta_{i_1i_2}}{c_2(c_2+D\bar c_2)}\mathcal{P}_{i_3i_3}\nonumber \\
&+\frac{c_2'\delta_{i_1i_2}}{c_2c_0}\left( \frac{\bar c_2 D}{c_2+D\bar c_2}-1\right)\mathcal{P}_0,
\end{flalign*}
\begin{flalign*}
&\xi_{i_1}\xi_{i_2}\xi_{i_3} =\frac{1}{c_3} \mathcal{P}_{i_1i_2i_3}-\frac{\bar c_3}{c_3[c_3+\bar c_3(D+2)]}\nonumber \\
&\cdot(\mathcal{P}_{i_1i_4i_4}\delta_{i_2i_3}+\mathcal{P}_{i_2i_4i_4}\delta_{i_1i_3} + \mathcal{P}_{i_3i_4i_4}\delta_{i_1i_2}) - \frac{c_3'}{c_3c_1}\nonumber \\
&\cdot\left[\frac{\bar c_3(D+2)}{c_3 + \bar c_3 (D+2)} +1 \right](\mathcal{P}_{i_1}\delta_{i_2i_3} + \mathcal{P}_{i_2}\delta_{i_1i_3} + \mathcal{P}_{i_3}\delta_{i_1i_2}).
\end{flalign*}
Thus, to recover the full momentum conservation equation we need to expand the EDF until \textit{third order} ($K=3$) in generalized polynomials, because this is the highest polynomial order that appear. For the calculation of $\pi_{i_1i_2}$ only the zeroth and second order expansion terms are non-zero and for $\pi_{i_1i_2i_3}$ only the first and third order terms are relevant.

Analogously, we find out that to recover the energy conservation equation, we need the \textit{fourth order} ($K=4$) expansion, since $\phi_{i_1i_2}$ in Eq.\eqref{phi2-eq} has a fourth order monomial. 
Hence we have shown that the macroscopic equations given in Sec. \ref{mac-eq-original-sec} stem from general arguments and so, are equations that govern the semiclassical fluids obtained with the original EDF since higher order terms in the series expansion do not contribute at all although they are there. In the LBM one takes advantage of this fact by eliminating the higher order terms. For this reason it is usual for practical purposes, to retain terms in the series expansion of the EDF only up to second order.
This simplifies the computational models, applicable to a small Mach numbers in case of classical particles. We exploit a similar model for the semiclassical particles.

\section{Forcing term}\label{forcing-sec}
In this section we treat the presence of  a forcing field in the Boltzmann-BGK equation describing a semiclassical fluid. The forcing term in the Boltzmann equation is given by  $\boldsymbol{a}\cdot \nabla _\xi f$ (see Eq.\eqref{boltz-cont}) and must satisfy the following moment constraints~\cite{PhysRevE.62.4982}:
\begin{flalign}
& \int d^D\xi \boldsymbol{a}\cdot \nabla_\xi f = 0 \label{force-eq1}\\
&  \int d^D\xi \boldsymbol{\xi}\boldsymbol{a}\cdot \nabla_\xi f = \int d^D \xi \boldsymbol{a} f  \label{force-eq2}\\
&  \int d^D\xi \xi_{i_1} \xi_{i_2} \boldsymbol{a}\cdot \nabla_\xi f = \int d^D \xi (\xi_{i_1} a_{i_2} + \xi_{i_2} a_{i_1})f. \label{force-eq3}
\end{flalign}
The moments of the forcing term up to second order (equations above) are the same for $f$ and $f^{eq}$ according to the Chapmann-Enskog assumption. If the force does not depend on $\xi$, we have the usual moments
\begin{flalign}
 &\int d^D \xi \boldsymbol{a} f   = -\rho \boldsymbol{a} \label{force-eq2-const} \:\:\:\mbox{and}  \\ &
\int d^D \xi (\xi_{i_1} a_{i_2} + \xi_{i_2} a_{i_1})f = -\rho (u_{i_1}a_{i_2} + u_{i_2}a_{i_1}). \label{force-eq3-const}
\end{flalign}
However for semiclassical fluids the particles can be  charged, such as in case of electrons in metals, a feature not commonly found in classical fluids. Therefore the Lorentz force must be included in its full account and so, there are two types of accelerations, one due to an electrical field $\boldsymbol{a_E}=\boldsymbol{E}$, which does not depend on the microscopic velocity, and another one, due to a magnetic field $\boldsymbol{a_B}=\boldsymbol{\xi}\times\boldsymbol{B}$, which does depend on the microscopic velocity. Recall that we have adopted natural units: $e=m_e=1$.

In practice, the forcing term can be easily implemented in LBM simulations by updating the macroscopic velocity in the EDF~\cite{sukop06,PhysRevE.53.743} as follows:
\begin{flalign}\label{u-update-force-term-eq}
\boldsymbol{u}_{t+\Delta t} = \boldsymbol{u}_{t}+ \tau \boldsymbol{a} ,
\end{flalign}
where $\boldsymbol{a}$ is the acceleration. This approach is equivalent to the previous one up to order $\mathcal{O}(\tau^2)$.

Therefore our goal in this section is to verify that these moment constraints are respected by the forcing term built with the semiclassical EDF, as similarly done in Ref.~\cite{PhysRevE.62.4982} for the classical EDF.
For the  sake of simplicity, we verify explicitly  the moment constraints only until second order, which means that we are considering the expanded EDF given by Eq.\eqref{edf-second-order-exp-eq}. Nevertheless we recall that such relations must hold to any expansion order. Since the EDF is a function of $(\boldsymbol{\xi}-\boldsymbol{u})$, we have that $  \boldsymbol{a}\cdot \nabla_\xi f^{eq} = - \boldsymbol{a}\cdot \nabla_u f^{eq} $. Therefore,
\begin{flalign*}
& - \boldsymbol{a} \cdot \nabla _u f^{eq}  = - \rho \omega(\xi) \Big[ c_1^2 \boldsymbol{\xi} + c_2^2 (\boldsymbol{\xi}\cdot \boldsymbol{u}) \boldsymbol{\xi} + (2c_2 \bar c_2 \nonumber \\ &
 + D \bar c_2^2)\xi^2 \boldsymbol{u} + (c_2c_2' + D \bar c_2 c_2')\boldsymbol{u} \Big ] \cdot \boldsymbol{a} ,
\end{flalign*}
or, using the coefficient definitions, we can write it in terms of the integrals:
\begin{flalign}\label{forcing-term-integrals-eq}
& - \boldsymbol{a} \cdot \nabla _u f^{eq}  = - \rho \omega(\xi) \Big[  \frac{1}{I_2}\boldsymbol{\xi} + \frac{1}{I_4}(\boldsymbol{\xi}\cdot \boldsymbol{u})\boldsymbol{\xi} \nonumber \\ &
+ \frac{(\Delta_2^2 -1)}{I_4 D}\xi^2 \boldsymbol{u} - \frac{I_2}{I_0 I_4} \Delta_2^2\boldsymbol{u}  \Big ] \cdot \boldsymbol{a} ,
\end{flalign}
We proceed with the demonstration that Eqs. \eqref{force-eq1}, \eqref{force-eq2} and \eqref{force-eq3} hold for the
EDF given in Eq.\eqref{edf-second-order-exp-eq}.\\

\noindent $\bullet$ {\bf Eq.\eqref{force-eq1}} -- Considering $\boldsymbol{a_E}$ first, the Eq.\eqref{forcing-term-integrals-eq} replaced in Eq.\eqref{force-eq1} gives:
\begin{flalign*}
&\int d^D \xi(- \boldsymbol{a_E}\cdot \nabla_u f^{eq}) = -\int d^D\xi \rho\omega(\xi)\Big[ \frac{1}{I_2}\xi_{i_1} E_{i_1} \\ &
+ \frac{1}{I_4} \xi_{i_3} u_{i_3} \xi_{i_1} E_{i_1} + \frac{(\Delta_2^2-1)}{I_4D}\xi_{i_3}\xi_{i_3}u_{i_1}E_{i_1} \\ &- \frac{I_2}{I_0I_4}\Delta_2^2u_{i_1}E_{i_1}\Big]
=-\rho\Big[\frac{1}{I_4}u_{i_3}E_{i_1}I_2\delta_{i_3i_1}\\ &+ \frac{(\Delta_2^2-1)}{I_4D}u_{i_1}E_{i_1}I_2D  - \frac{I_2}{I_4I_0}\Delta_2^2u_{i_1}E_{i_1}I_0 \Big]
=0,
\end{flalign*}
where we have used the definitions of the integrals, Eq.\eqref{i2n-cont}, and the fact that odd powers of $\xi$ give null integrals. Thus, for a force that does not depend on $\xi$, the Eq. \eqref{force-eq1} is demonstrated. Now, for a magnetic acceleration $\boldsymbol{a_B}=\boldsymbol{\xi}\times \boldsymbol{B} \Rightarrow (a_B)_{i_1}=\epsilon _{i_1i_2i_3} \xi_{i_2} B_{i_3}$, where $\epsilon _{i_1i_2i_3}$ is the Levi-Civita tensor, we have:
\begin{flalign*}
&\int d^D \xi (-\boldsymbol{a_B}\cdot \nabla _u f^{eq}) \\ &
= -\int d^D \xi \omega (\xi)\rho \Big [ \frac{1}{I_2} \xi_{i_1} \xi_{i_2} B_{i_3} \epsilon_{i_1i_2i_3} \\ &
+ \frac{1}{I_4} \xi_{i_4} u_{i_4} \xi_{i_1} \xi_{i_2} B_{i_3} \epsilon_{i_1i_2i_3} + \frac{(\Delta_2^2-1)}{I_4D}\xi_{i_4}\xi_{i_4} u_{i_1}\xi_{i_2}B_{i_3}\epsilon_{i_1i_2i_3}  \\ &
-\frac{I_2}{I_0I_4}\Delta_2^2 u_{i_1}\xi_{i_2} B_{i_3} \epsilon_{i_1i_2i_3}   \Big]
=-  \frac{\rho}{I_2}B_{i_3}I_2\delta_{i_1i_2}\epsilon_{i_1i_2i_3} =0
\end{flalign*}
So, Eq. \eqref{force-eq1} is demonstrated for the two cases.

\noindent $\bullet$ {\bf Eq.\eqref{force-eq2}} -- For $\boldsymbol{a_E}$:
\begin{flalign*}
&\int d^D \xi \xi_{i_4} (-\boldsymbol{a_E}\cdot \nabla_uf^{eq}) = - \int d^D \xi \omega(\xi)\rho \\ &
\cdot\Big[ \frac{1}{I_2}\xi_{i_1}\xi_{i_4} E_{i_1} + \frac{1}{I_4}\xi_{i_3} u_{i_3} \xi_{i_1} E_{i_1}\xi_{i_4} \\ &
+ \frac{(\Delta_2^2 -1)}{I_4D}\xi_{i_3}\xi_{i_3}u_{i_1}E_{i_1}\xi_{i_4} - \frac{I_2}{I_4} \Delta_2^2 u_{i_1} E_{i_1} \xi_{i_4} \Big] \\ &
=\frac{\rho}{I_2}E_{i_1}I_2 \delta_{i_1i_4} = -\rho (a_E)_{i_4}.
\end{flalign*}
For $\boldsymbol{a_B}$:
\begin{flalign*}
 &\int d^D \xi \xi_{i_5} (-\boldsymbol{a_B}\cdot \nabla_u f^{eq})= -\int d^D\xi \omega(\xi) \rho  \\ &
 \cdot\Big[   \frac{1}{I_2}\xi_{i_1}\xi_{i_2}\xi_{i_5} B_{i_3} \epsilon_{i_1i_2i_3} + \frac{1}{I_4}\xi_{i_4}u_{i_4}\xi_{i_1} \xi_{i_2} \xi_{i_5} B_{i_3} \epsilon_{i_1i_2i_3}   \\ &
 - \frac{I_2}{I_0I_4}\Delta_2^2 u_{i_1} \xi_{i_2} \xi_{i_5} B_{i_3} \epsilon_{i_1i_2i_3} \\&+ \frac{(\Delta_2^2-1)}{I_4D}\xi_{i_4} \xi_{i_4} u_{i_1} \xi_{i_2} B_{i_3} \xi_{i_5} \epsilon_{i_1i_2i_3}   \Big] \\&
 = - \rho \Big[ \frac{1}{I_4}u_{i_4}B_{i_3}\epsilon_{i_1i_2i_3}I_4\delta_{i_1i_2i_4i_5} - \frac{I_2}{I_4}\Delta_2^2 B_{i_3} u_{i_1} \epsilon_{i_1i_2i_3}\\ & \cdot I_2 \delta_{i_2i_5}  + \frac{(\Delta_2^2-1)}{I_4D}u_{i_1}B_{i_3}I_4\delta_{i_4i_4i_2i_5}\epsilon_{i_1i_2i_3}   \Big]
\end{flalign*}
but $\epsilon_{i_1i_2i_3}\delta_{i_1i_2i_4i_5} = 0$ and $\epsilon_{i_1i_2i_3}\delta_{i_4i_4i_2i_5}=-(D+2)\epsilon_{i_1i_3i_5}$, so:
\begin{flalign*}
 &\int d^D \xi \xi_{i_5} (-\boldsymbol{a_B}\cdot \nabla_u f^{eq})=   \rho \Big[ -\frac{I_2^2}{I_0I_4}\Delta_2^2 B_{i_3} u_{i_1} \epsilon_{i_1i_3i_5} \\ &
 + \frac{1}{D}(\Delta_2^2-1)(D+2)u_{i_1}B_{i_3} \epsilon_{i_1i_3i_5} \Big] \\ &
 = - \rho \Big [ J_2 \Delta_2^2 - \frac{D+2}{D}(\Delta_2^2 - 1) \Big]  (\boldsymbol{u}\times \boldsymbol{B})_{i_5} \\ &
 = - \rho (a_B)_{i_5} ,
\end{flalign*}
where we used the identity $J_2\Delta_2^2 - (D+2)/D(\Delta_2^2 -1) = 1$ that can be shown with the expressions for $J_2$ and $\Delta_2$ in terms of the integrals. Thus, the Eq.\eqref{force-eq2}, which was equivalent to Eq.\eqref{force-eq2-const}, is also verified.

\noindent $\bullet$ {\bf Eq.\eqref{force-eq3}} -- For $\boldsymbol{a_E}$,
\begin{flalign*}
 &\int d^D \xi \xi_{i_5} \xi_{i_6} (-\boldsymbol{a_E}\cdot \nabla_u f) = - \int d^D \xi \omega(\xi) \rho \\&
 \cdot \Big[ \frac{1}{I_2}\xi_{i_1}\xi_{i_5} \xi_{i_6} E_{i_1} + \frac{1}{I_4} \xi_{i_3} u_{i_3} \xi_{i_1} E_{i_1} \xi_{i_5} \xi_{i_6} \\ &
 + \frac{(\Delta_2^2-1)}{I_4D} \xi_{i_3} \xi_{i_3} u_{i_1} E_{i_1} \xi_{i_5} \xi_{i_6}
 - \frac{I_2}{I_0I_4} \Delta_2^2 u_{i_1} E_{i_1} \xi_{i_5} \xi_{i_6} \Big] \\ &
 = -  \rho \Big[  \frac{1}{I_4}u_{i_3} E_{i_1} I_4 \delta_{i_3i_1i_5i_6}+\frac{I_4(\Delta_2^2-1)}{I_4D} u_{i_1} E_{i_1} \delta_{i_3i_3i_5i_6} \\&
 - \frac{I_2}{I_0I_4}\Delta_2^2u_{i_1}E_{i_1} I_2 \delta_{i_6i_5}  \Big]
 = -  \rho   (u_{i_5}E_{i_6}+u_{i_6}E_{i_5})  \\ &
 - \rho u_{i_1}E_{i_1}\delta_{i_5i_6}\Big[ 1+\frac{(\Delta_2^2 -1)}{D}(D+2) - J_2\Delta_2^2  \Big] \\ &
 = -  \rho   (u_{i_5}E_{i_6}+u_{i_6}E_{i_5}) ,
\end{flalign*}
giving the Eq.\eqref{force-eq3-const} as expected. For $\boldsymbol{a_B}$:
\begin{flalign*}
 &\int d^D \xi \xi_{i_5} \xi_{i_6} (-\boldsymbol{a_B}\cdot\nabla_u f^{eq}) = - \int d^D \xi \omega(\xi) \rho \\ & \cdot\Big[ \frac{1}{I_2}\xi_{i_1}\xi_{i_2}\xi_{i_5}\xi_{i_6}B_{i_3}\epsilon_{i_1i_2i_3} + \frac{1}{I_4}\xi_{i_4}u_{i_4}\xi_{i_1}\xi_{i_2}\xi_{i_5} \xi_{i_6} B_{i_3} \epsilon_{i_1i_2i_3} \\ &
 - \frac{I_2}{I_0I_4}\Delta_2^2u_{i_1}\xi_{i_2}\xi_{i_5}\xi_{i_6}B_{i_3}\epsilon_{i_1i_2i_3} - \frac{I_2}{I_0I_4}\Delta_2^2 u_{i_1}\xi_{i_2}\xi_{i_5}\xi_{i_6}  \\ & \cdot B_{i_3}\epsilon_{i_1i_2i_3}   +\frac{(\Delta_2^2-1)}{I_4D}\xi_{i_4}\xi_{i_4}u_{i_1}\xi_{i_2}B_{i_3}\xi_{i_5}\xi_{i_6}\epsilon_{i_1i_2i_3}    \Big]   \\ &
 = -  \rho \frac{I_4}{I_2} \delta_{i_1i_2i_5i_6} B_{i_3}\epsilon_{i_1i_2i_3} = 0
\end{flalign*}
which is the expected result since
\begin{flalign*}
 &\int d^D \xi (\xi_{i_5} (a_B)_{i_6} + \xi_{i_6} (a_B)_{i_5})f^{eq} \\ &= \int d^D \xi (\xi_{i_5} \epsilon_{i_6i_7i_8}\xi_{i_7} B_{i_8} + \xi_{i_6}\epsilon_{i_5i_7i_8}\xi_{i_7}B_{i_8})f^{eq} \\&
 = B_{i_8} \epsilon_{i_6i_7i_8} \frac{1}{D}\int d^D \xi \xi^2 f^{eq} \delta_{i_5i_7} \\ &+ B_{i_8} \epsilon_{i_5i_7i_8}\frac{1}{D}\int d^D \xi \xi^2 f^{eq} \delta_{i_6i_7} =0
\end{flalign*}
Note that, for $\boldsymbol{a_B}$, the equations \eqref{force-eq3} and \eqref{force-eq3-const} are not equivalent.\\

In summary, we have shown that the constraints of Eqs. \eqref{force-eq1}, \eqref{force-eq2} and \eqref{force-eq3} are satisfied for a forcing term calculated explicitly with the EDF given in Eq.\eqref{edf-second-order-exp-eq}.

\section{Quadrature beyond Gauss-Hermite}\label{quadrature-sec}

In this section, we extend the concept of quadrature beyond the Gauss-Hermite concept~\cite{abramowitz1964handbook}, which means that the weight function $\omega(\xi)$ is not necessarily the gaussian function well suited for the D-dimensional Hermite polynomials (Eq.\eqref{hermite-weight-eq}). Here we are interested in the generalized polynomials applicable to the semiclassical LBM.
The basic assumption is that there is a discrete space of microscopic velocities $\xi_\alpha$ where integrals can be replaced by sums where a discrete set of weights $w_\alpha$ play the role of the weight function $\omega(\xi)$.
The following equations should be satisfied where $I_M$ are known before hand from Eq.\eqref{i2n-cont}:
\begin{flalign}\label{quadrature-eq}
 \sum_\alpha w_\alpha \xi_{\alpha i_1}\xi_{\alpha i_2}\ldots \xi_{\alpha i_M} &= \int d^D \xi \omega(\xi) \xi_{i_1} \xi_{i_2} \ldots \xi_{i_M},
\end{flalign}
Hence we demand that the integral of Eq.\eqref{quadrature-eq} be equal to $I_M \delta_{i_1i_2\ldots i_M}$, according to  Eq.\eqref{i2n-cont}, to obtain that,
\begin{flalign}\label{quadrature-eq-2}
 \sum_\alpha w_\alpha \xi_{\alpha i_1}\xi_{\alpha i_2}\ldots \xi_{\alpha i_M} &= I_M \delta_{i_1i_2\ldots i_M}.
\end{flalign}
Notice that for the Gauss-Hermite quadrature $I_M=1$ and $J_M=1$ but not for a general weight function.
One of the key and well-known features of the LBM is that it takes advantage that only a few of such conditions have to be implemented in order to reach the conservation of mass, momentum and energy.
This gives rise to the discrete lattices where only a finite set of discrete weights $w_\alpha$ is obtained that solve the above relations up to a maximal $M$.
Next we determine some of these sets to be used with the semiclassical Boltzmann-BGK equation.
We use the standard nomenclature ``DdVv'', where ``d'' denotes the dimension and ``v'' the number of lattice vectors.
We have defined the lattice vectors $e_\alpha$ proportional to the discrete velocities $\xi_\alpha$, such that $\boldsymbol{\xi_\alpha} = \boldsymbol{e_\alpha} /c_s$, where $c_s$ is the reference speed to be found by solving the quadrature equations. By introducing the reference speed one can choose to define one of the lattice vectors, usually the one oriented along the positive x axis, to be equal to one. Below we see the first six quadrature equations:
\begin{flalign*}
&\sum_\alpha w_\alpha  = I_0 ,\\ &
\sum_\alpha w_\alpha e_{\alpha i_1} = 0,\\&
\sum_\alpha w_\alpha e_{\alpha i_1}e_{\alpha i_2} = I_2 c_s^{2} \delta_{i_1i_2} ,\\ &
\sum_\alpha w_\alpha e_{\alpha i_1}e_{\alpha i_2}e_{\alpha i_3} = 0, \\&
\sum_\alpha w_\alpha e_{\alpha i_1}e_{\alpha i_2}e_{\alpha i_3}e_{\alpha i_4} = I_4 c_s^{4} \delta_{i_1i_2i_3i_4},\\&
\sum_\alpha w_\alpha e_{\alpha i_1}e_{\alpha i_2}e_{\alpha i_3}e_{\alpha i_4}e_{\alpha i_5} = 0,\:\:\:\:
(\ldots).
\end{flalign*}
The order $M$ (see Eq.\eqref{quadrature-eq}) is an important characteristic of the quadrature, since it gives the maximum moment of the weight function for which the quadrature provides equivalence between sums and integrals. Notice that the quadrature equations with $M$ odd are automatically satisfied due to the symmetry among vectors $e_\alpha$, and so, give trivial expressions. In table \ref{lattices-tab} we see the order $M$ of some quadratures. It should be noticed that all $w_\alpha$ and $c_s$ must be positive quantities in order to have stable simulations.
Next, we explicitly calculate one quadrature for each dimension: D1V3, D2V9 and D3V15. More quadratures can be found in \ref{quadrature-appendix}.

\subsection{D1V3}

The lattice vectors for this lattice are $ e_\alpha = {-1, \, 0, \,+1}$ and there are two different weight: $w_0$ for $e_0 = 0$ and $w_1$ for $e_\pm = \pm 1$. So we have three variables to be determined: the two weights and the reference speed $c_s$. We need to solve three quadrature equations:
\begin{flalign*}
&\sum_\alpha w_\alpha = I_0 \Rightarrow w_0 + 2w_1 = I_0,\\
&\sum_\alpha w_\alpha e_\alpha^2 = I_2 c_s^{2} \Rightarrow 2w_1 = I_2 c_s^{2},\\
&\sum_\alpha w_\alpha e_\alpha^4 = 3I_4 c_s^{4} \Rightarrow 2w_1 = 3I_4 c_s^{4}.
\end{flalign*}
The solution for this system is:
\begin{flalign*}
c_s = \sqrt{\frac{I_2}{3I_4}}, \:\:\: w_0 = I_0\left( 1-\frac{J_2}{3}\right),\: \:\: w_1 = \frac{I_0 J_2}{6}.
\end{flalign*}
For the Hermite weight function, this solution becomes the standard D1V3 lattice: $w_0=2/3$, $w_1=1/6$ and $c_s=1/\sqrt{3}$. For the D1V3 lattice, $M=5$, meaning that moments up to order five in Eq.\eqref{quadrature-eq} are exactly calculated by the sums.

\subsection{D2V9}

The lattice vectors are $e_s=[(\pm 1,0),(0,\pm 1)]$, $e_l=[(\pm 1, \pm 1)]$ and $e_0 = (0,0)$, with weights $w_s$, $w_l$ and $w_0$ respectively. As we have four unknowns (three weights and $c_s$), we need four equations, which are:

\begin{flalign*}
\sum_\alpha w_\alpha& = I_0 \Rightarrow w_0+4w_s + 4w_l = I_0 \\
\sum_\alpha w_\alpha& e_{\alpha i_1} e_{\alpha i_2} = I_2 c_s^{2}\delta_{i_1i_2} \Rightarrow 2w_s + 4w_l = I_2 c_s^{2}\\
\sum_\alpha w_\alpha& e_{\alpha i_1} e_{\alpha i_2} e_{\alpha i_3} e_{\alpha i_4} = I_4c_s^{4} \delta_{i_1i_2i_3i_4}
\\ &\Rightarrow
\begin{cases}
      2w_s + 4w_l = 3I_4 c_s^{4} \\
      4w_l = I_4 c_s^{4}
\end{cases}
\end{flalign*}
where the last quadrature equation split in two equations because there are two possible choices for the indexes that give non-trivial equations: one for $i_1=i_2=i_3=i_4$ and other for $(i_1=i_2)\neq (i_3=i_4)$. The solution is:
\begin{flalign*}
&w_0 = I_0 \left( 1-\frac{5J_2}{9}\right), \,\,\, w_s = \frac{I_0J_2}{9},\nonumber \\
& w_l = \frac{I_0 J_2}{36}, \,\,\, c_s = \sqrt{\frac{I_2}{3I_4}}.
\end{flalign*}
It gives the standard D2V9 for the Hermite weight: $w_0=4/9$, $w_s = 1/9$, $w_l =1/36 $ and $c_s = 1/\sqrt{3}$.

\subsection{D3V15}

The lattice vectors are $e_0 = (0,0,0)$, $e_s=[(\pm 1,0,0),(0,\pm 1,0), (0,0,\pm 1)]$, $e_l=[(\pm 1, \pm 1, \pm 1)]$ with respective weights $w_0$, $w_s$ and $w_l$. Quadrature equations:
\begin{flalign*}
\sum_\alpha w_\alpha &= I_0 \Rightarrow w_0+6w_s + 8w_l = I_0\\
\sum_\alpha w_\alpha &e_{\alpha i_1} e_{\alpha i_2} = I_2 c_s^{2}\delta_{i_1i_2} \Rightarrow 2w_s + 8w_l = I_2 c_s^{2}\\
\sum_\alpha w_\alpha &e_{\alpha i_1} e_{\alpha i_2} e_{\alpha i_3} e_{\alpha i_4} = I_4c_s^{4} \delta_{i_1i_2i_3i_4}\\ &\Rightarrow
\begin{cases}
      2w_s + 8w_l = 3I_4 c_s^{4} \\
      8w_l = I_4 c_s^{4}
\end{cases}
\end{flalign*}
Solutions:
\begin{flalign*}
&w_0 = I_0 \left( 1-\frac{7J_2}{9}\right), \,\,\, w_s = \frac{I_0J_2}{9},\nonumber \\
& w_l = \frac{I_0 J_2}{72}, \,\,\, c_s = \sqrt{\frac{I_2}{3I_4}}.
\end{flalign*}
It gives the standard D3V15 for the Hermite weight: $w_0=2/9$, $w_s = 1/9$, $w_l = 1/72$ and $c_s = 1/\sqrt{3}$

\section{The isothermal lattice Boltzmann method for electrons in metals}\label{lbm-electrons-sec}
\subsection{Model description}\label{model-sec}

In this section, we build a simple and efficient model for electrons in metals in 2D and 3D dimensions and test it with the Riemann problem, the Poiseuille flow and the Ohm's law.
The model complies with the condition $\omega(\xi) \approx f^{eq}(\xi)$, thus the weight is not equal to the EDF itself. The electrons inside the metal are governed by the Fermi-Dirac distribution~\cite{ibach03},
\begin{flalign*}
 f^{eq}_{FD} = \left\{\exp \left[ \frac{m_e(\boldsymbol{\chi}-\boldsymbol{v})^2}{2k_BT}-\frac{\mu'}{k_BT}  \right]+1\right\}^{-1} ,
\end{flalign*}
where $\chi$ and $v$ are the microscopic and macroscopic velocities respectively, $m_e$ is the electron mass, $k_B$ is the Boltzmann constant, $T$ is the temperature and $\mu'$ is the chemical potential. The Fermi energy can be expressed as a function of the Fermi temperature $T_F$ and the Fermi speed $v_F$ as
\begin{flalign*}
E_F=k_B T_F = \frac{1}{2}m_e v_F^2.
\end{flalign*}
Considering the Fermi speed, $v_F=\sqrt{2k_B T_F/m_e}$, as the reference speed for our model, we can define non-dimensional variables:
\begin{flalign*}
\mu \equiv \frac{\mu'}{k_B T_F}, \:\:\:\:\theta \equiv \frac{T}{T_F}, \:\:\:\:\boldsymbol{\xi}\equiv\frac{\boldsymbol{\chi}}{v_F} \:\:\: \:\mbox{and}\:\:\:\: \boldsymbol{u} \equiv \frac{\boldsymbol{v}}{v_F}
\end{flalign*}
leading to
\begin{flalign}
f^{eq}_{FD} = \left\{\exp \left[ \frac{(\boldsymbol{\xi}-\boldsymbol{u})^2}{\theta}-\frac{\mu}{\theta}  \right]+1\right\}^{-1}.
\label{fd-dist-non-dim-eq}
\end{flalign}
We consider in our model the physical parameters of cooper at room temperature ($T=300$ K), which has Fermi temperature $T_F^{Cu}=8.16\times 10^4\, K$, giving $\theta \approx 1/270$, $\mu=1$ and $z=e^{270}$~\cite{ashcroft76}. We expand the Fermi-Dirac distribution in Eq.\eqref{fd-dist-non-dim-eq} up to second order in generalized polynomials, where the coefficients are calculated by orthogonalizing the polynomials with respect to the weight function below:
\begin{flalign}
\omega(\xi)=\frac{1}{e^{-270}e^{270\xi^2}+1},
\label{weight-func-isotherm-eq}
\end{flalign}
which, initially, is the Eq.\eqref{fd-dist-non-dim-eq} for $\boldsymbol{u}=0$ and constant $\theta$ and $\mu$. Hence the condition $\omega(\xi) = f^{eq}(\xi)$ discussed along this paper is not being implemented here otherwise the weights would have to be updated at each time step since the chemical potential changes in $f^{eq}(\xi)$ while here $\omega(\xi)$ remains constant. The integrals $I_N$, Eq.\eqref{i2n-cont2}, are different for 2D and 3D, which implies that the polynomial coefficients are also different for the two cases (see Sec. \ref{polynomials-sec}). The discrete version of the second order expansion in Eq.\eqref{edf-second-order-exp-eq} becomes:
\begin{flalign}
 &f^{eq}_\alpha = \rho \,w_\alpha \Big\{  c_0^2 + c_1^2(\boldsymbol{\xi_\alpha}\cdot\boldsymbol{u}) + \frac{1}{2}c_2(c_2\bar \theta+ c_2')\xi_\alpha^2 \nonumber\\ &
 + \frac{c_2^2}{2}(\boldsymbol{\xi_\alpha}\cdot\boldsymbol{u})^2 + \frac{1}{2}c_2 \bar c_2 (D\bar \theta + u^2)\xi_\alpha^2+\frac{1}{2}(\bar c_2\xi_\alpha^2 + c_2')\nonumber \\ &
 \cdot [D(c_2\bar\theta+ c_2')+c_2 u^2+D\bar c_2 (D\bar \theta+u^2)] \Big\} ,
 \label{feq-exp-2nd-order-eq}
\end{flalign}
where $\bar \theta$ have different values for 2D and 3D and $\boldsymbol{\xi_\alpha}=\boldsymbol{e_\alpha}/c_s$ are the discrete velocities given in Sec. \ref{quadrature-sec}. Besides having constant temperature $\theta=1/270$ we use another approximation for the semiclassical model in order to simplify the numerical implementation, which is $\bar\theta  = I_2/I_0 = \theta g_{\frac{D}{2}+1}(z) /(2g_{\frac{D}{2}}(z))= \mbox{constant}$. For the classical LBM, $\bar\theta$ is automatically constant for the isothermal case, since $\bar \theta_{cl} = \theta$. Therefore, this extra approximation is not needed. With this approximations and using the expressions for the coefficients, we verify the identity $c_2\bar\theta+c_2'+D\bar c_2 \bar \theta = 0$, which leads to:
\begin{flalign}\label{feq-reduced-applied-electrons-in-metals-eq}
 &f^{eq}_\alpha = \rho \,w_\alpha \Big\{  c_0^2 + c_1^2(\boldsymbol{\xi_\alpha}\cdot\boldsymbol{u})
 + \frac{c_2^2}{2}(\boldsymbol{\xi_\alpha}\cdot\boldsymbol{u})^2 \nonumber\\ &+ \frac{1}{2}c_2 \bar c_2 u^2\xi_\alpha^2+\frac{1}{2}(\bar c_2\xi_\alpha^2 + c_2')(c_2+D\bar c_2)u^2 \Big\} .
\end{flalign}
The above EDF is the one used in our numerical algorithm together added to the values for the coefficients and $\bar\theta$ given in the next sections for $D=2 \,\mbox{or}\,3$ dimensions, respectively. The quadratures are given in the next sections and the time evolution is governed by the Boltzmann equation in its discrete form (see Eq.\eqref{boltz-disc-eq}) and in terms of the lattice vectors $e_\alpha$.
\begin{flalign*}
&f_\alpha(\mathbf{x}+\mathbf{e}_\alpha\Delta t, t+\Delta t) - f_\alpha(\mathbf{x},t) \nonumber \\ &
= - \frac{\Delta t }{\tau} [f_\alpha (\mathbf{x},t)-f_\alpha^{eq}(\mathbf{x},t)],
\end{flalign*}
The macroscopic quantities are calculated by:
\begin{flalign}
\rho = \sum_\alpha w_\alpha f_\alpha , \:\:\:\: \:\mathbf{u}=\frac{1}{\rho}\sum_\alpha w_\alpha f_\alpha \frac{\boldsymbol{e_\alpha}}{c_s} .
\label{mac-fields-model-sec-eq}
\end{flalign}
We can convert the density into the chemical potential and vice-versa by means of the relation (see Eq.\eqref{rho-def-eq})
\begin{flalign*}
\rho = (\pi\theta)^{D/2} g_{\frac{D}{2}}(e^{\frac{\mu}{\theta}}).
\end{flalign*}
For $\mu/\theta \gg 1$, one can use the Sommerfeld expansion to approximate the FD integral:
\begin{flalign}\label{sommerfeld-approx-eq}
&g_\nu(e^{\frac{\mu}{\theta}}) = \frac{(\mu/\theta)^\nu}{\Gamma(\nu+1)}\Big[  1+\nu (\nu-1)\frac{\pi^2}{6}\left(  \frac{\theta}{\mu}\right)^{2} \nonumber \\ &
+ \nu (\nu-1)(\nu-2)(\nu-3)\frac{7\pi^4}{360}\left( \frac{\theta}{\mu} \right)^4 + \ldots  \Big]
\end{flalign}

\subsubsection{2D model}\label{2d-sec}

To build our model in 2D, we first calculate the polynomial coefficients using the weight function, Eq.\eqref{weight-func-isotherm-eq}, through their expressions given in Sec. \ref{polynomials-sec}.
\begin{center}
\begin{tabular}{ |l|l| }
 \hline
$c_0$ &   0.564189583547756286948079\\
$c_1$ &   1.128353706923879405456370\\
$c_2$ &   2.763766115146273701436833\\
$\bar c_2$ &  0.572262450908341120084001\\
$c_2'$ & -0.977116848075011682697851\\
$\bar \theta$ &0.250011282126658766985161\\
 \hline
\end{tabular}
\end{center}
where the pseudo-temperature was calculated by $\bar\theta  = I_2/I_0= \theta g_2(z) /(2g_1(z)) $. The discrete weights are also calculated using the weight function in Eq.\eqref{weight-func-isotherm-eq} to calculate the expressions in Table \ref{lattices-tab}. For a D2V9 lattice, we have:
\begin{center}
\begin{tabular}{ |l|l| }
 \hline
$w_0$ & 0.523716900428241365084608\\
$w_s$ & 0.523575150632310374675607\\
$w_l$ & 0.130893787658077593668902\\
$c_s$ & 1.414149748226522446289974\\
 \hline
\end{tabular}
\end{center}
Note that, although we are keeping the standard notation for the reference speed, $c_s$, it denotes the Fermi speed and not the sound speed as for the classical models.

To obtain the chemical potential from the density, we use the Sommerfeld's expansion, Eq.\eqref{sommerfeld-approx-eq}, leading to  $\rho = \pi \theta g_1(z) \approx \pi \mu$. This is an excellent approximation for the 2D case with accuracy much beyond the double precision ($10^{-16}$).
\begin{figure}[htb]
\center
\includegraphics[width=1.0\linewidth]{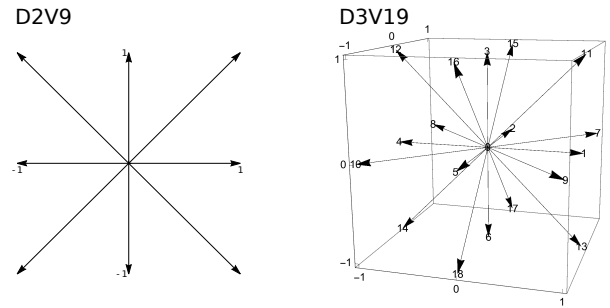}
\caption{Quadratures used in the models.}
\label{lattices-fig}
\end{figure}

\subsubsection{3D model}\label{3d-sec}

Following the same procedure as for the 2D model, we calculate the polynomial coefficients and $\bar\theta = \theta g_{\frac{5}{2}}(z)/(2g_{\frac{3}{2}}(z))$:
\begin{center}
\begin{tabular}{ |l|l| }
 \hline
$c_0$ &   0.488598377549843819982207\\
$c_1$ &   1.092502210196163024710861\\
$c_2$ &   2.890326124370599833053459\\
$\bar c_2$ &  0.559713196101887209686884\\
$c_2'$ & -0.913955004948841398767998  \\
$\bar\theta$ & 0.200013538215948856423209\\
 \hline
\end{tabular}
\end{center}
The discrete weights can be seem below:
\begin{center}
\begin{tabular}{ |l|l| }
 \hline
$w_0$ & 0.279433800596370971795231\\
$w_s$ & 0.325785607726861097214977\\
$w_l$ & 0.162892803863430548607489\\
$c_s$ & 1.527439075525116330156203\\
 \hline
\end{tabular}
\end{center}
The density can be calculated as $\rho = (\pi\theta)^{3/2} g_{\frac{3}{2}}(e^{\frac{\mu}{\theta}})$, where $\theta=1/270$ in our problem or, to extract the chemical potential from the density,
\begin{flalign*}
\mu = \theta \log \left[   g_{\frac{3}{2} }^{-1} \left(  \frac{\rho}{(\pi\theta)^{3/2}} \right)\right].
\end{flalign*}
Note that these relations between $\rho$ and $\mu$ are used just to set the initial conditions and to calculate the fields in the output. During the simulations just the density field is used. The Sommerfeld expansion can also be used but the accuracy is not as good as for the 2D case. The approximation with Eq.\eqref{sommerfeld-approx-eq} is reliable just up to $10^{-8}$, which is bellow the double precision. For this reason, we use numerical methods~\cite{press2015numerical} to calculate the FD integral and its inverse in order to obtain better accuracy.

\subsection{Riemann problem}\label{riemann-sec}

\begin{figure}[ht]
\center
\includegraphics[width=1.00\linewidth]{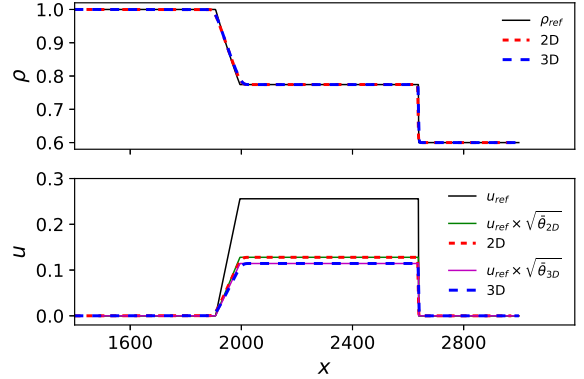}
\caption{(Color online) Solutions for the density and velocity fields in the Riemann problem obtained with our two models (2D and 3D) and with the analytical solution for the classical case. The classical solution for the velocity is corrected by $\sqrt{\bar\theta}$ in order to compare with the semiclassical results.}
\label{riemann-fig}
\end{figure}

The Riemann problem (or shock tube test) is a benchmark validation for computational fluid dynamic models and it consists in analyzing the shock waves formed when a discontinuity in the initial conditions evolves. This problem has analytical solutions for the inviscid case~\cite{toro09}. We simulate the Riemann problem using the two numerical methods (2D and 3D) in a effectively one-dimensional system: $L_X\times L_Y = 3000\times 2$ for the 2D model and $L_X\times L_Y \times L_Z = 3000 \times 2 \times 2$ for the 3D model. Initially, the density and velocity fields are the same for both models: $\rho=1.0$ inside the domain $L_X/4 < x <3L_X/4$ and $\rho=0.6$ outside and $\boldsymbol{u}=0$ everywhere. The relaxation time has the constant value $\tau=0.8$. The boundary conditions are periodic for all directions. In Fig. \ref{riemann-fig} we see the solutions given by our models and the analytical solutions for the classical case~\cite{toro09}. There is a good agreement for the densities field. For a sake of comparison, we correct the velocities give by the classical case by multiplying then by $\sqrt{\bar\theta}$, since $c_s \propto \sqrt{T}$ and $\bar\theta$ plays a role of an ``effective temperature'' for the semiclassical models. After this correction, the velocities also matches.

\subsection{Poiseuille flow}\label{poiseuille-sec}

\begin{figure}[ht]
\center
\includegraphics[width=1.0\linewidth]{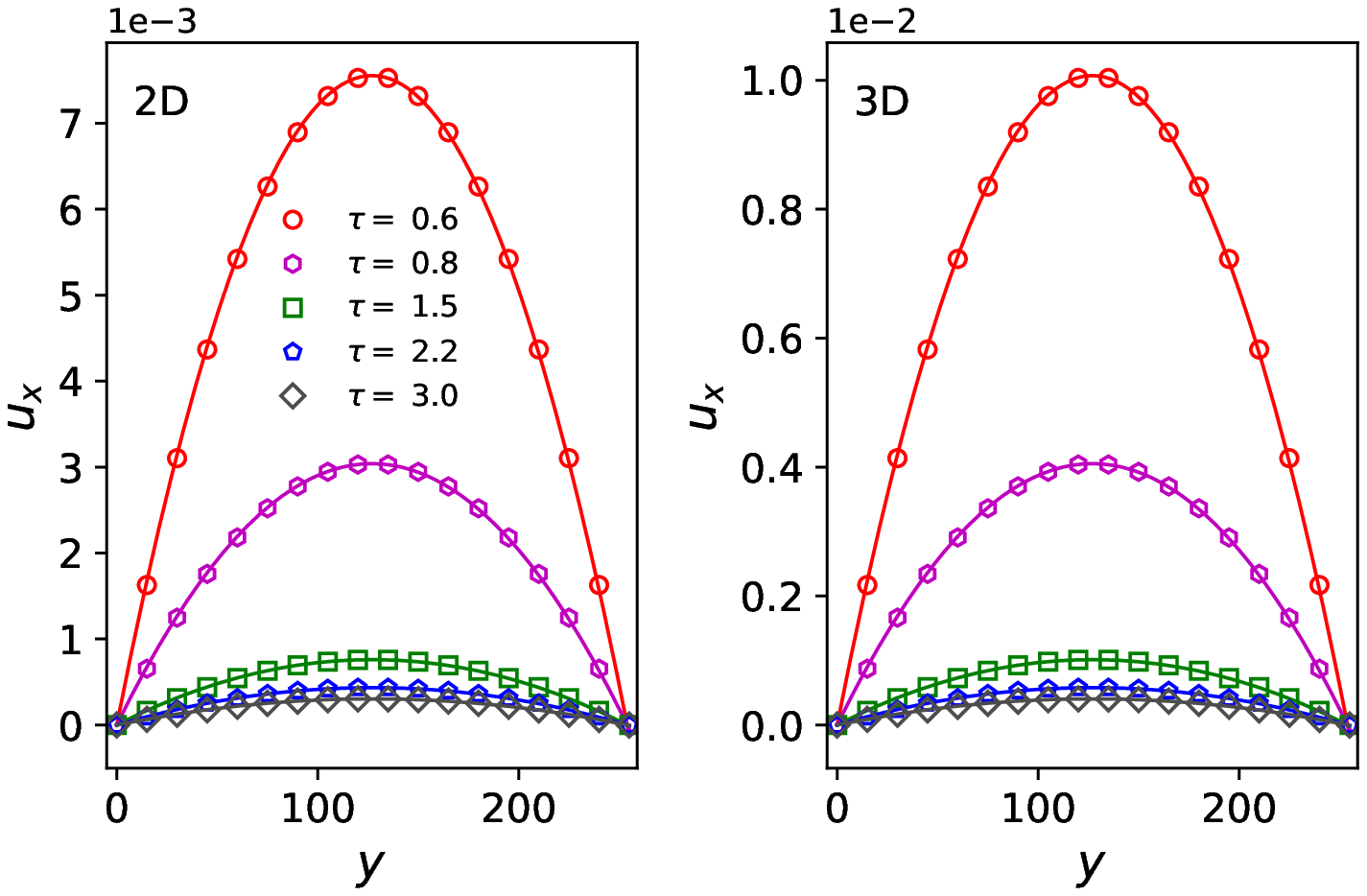}
\includegraphics[width=1.0\linewidth]{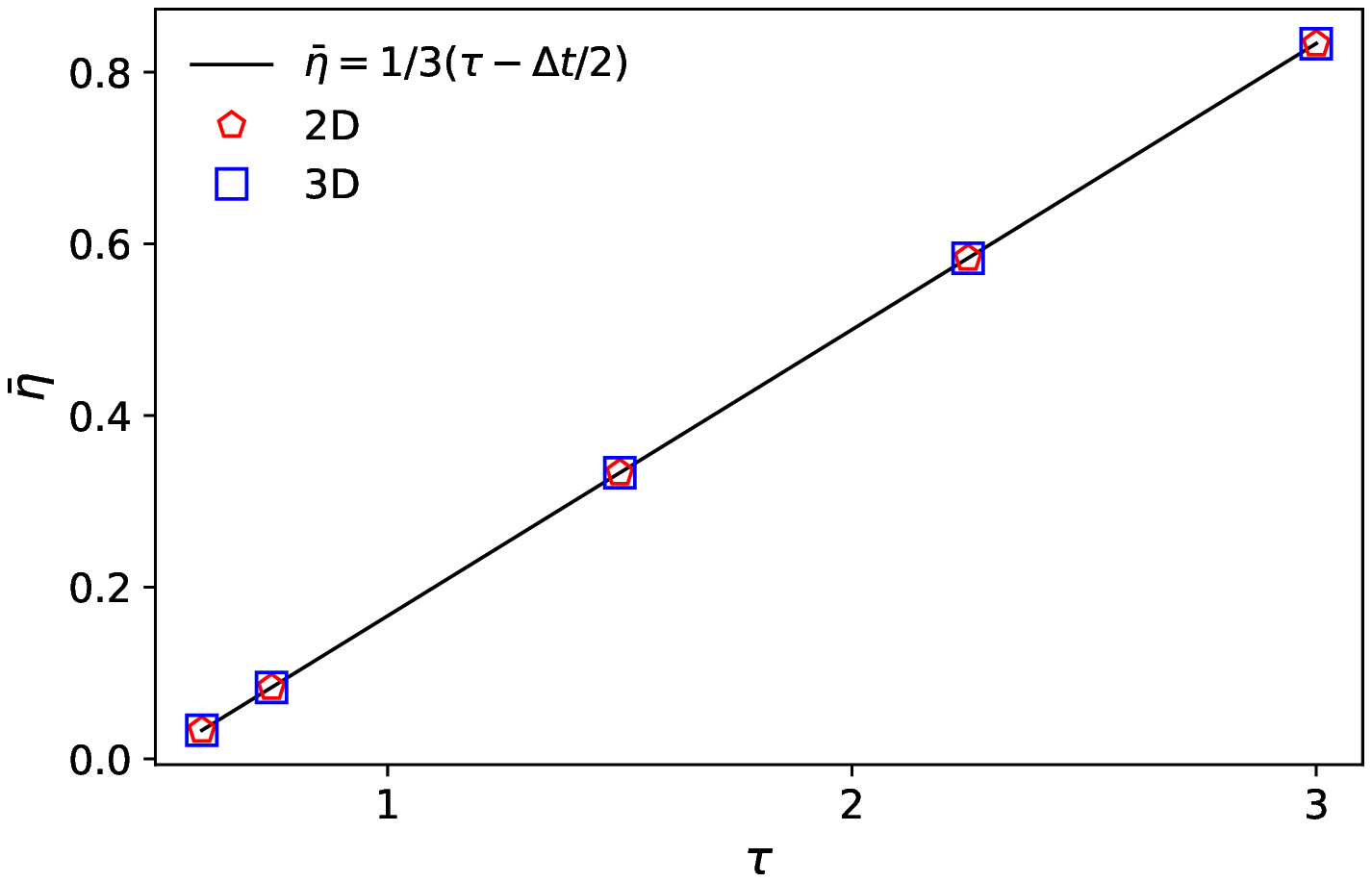}
\caption{(Color online) Velocity profiles for the Poiseuille flow obtained with the two models for five different relaxation times and the viscosity measurement for the two models.}
\label{poiseuille-fig}
\end{figure}

We simulate the viscous fluid of electrons in metals passing through a channel of constant cross section (parallel plates) and analyze the velocity profile for the steady state. Assuming a incompressible fluid submitted to an external force with acceleration $\mathbf{a}=a\mathbf{i}$ in the $x$ direction, the Navier-Stokes equation for semiclassical fluids, Eq.\eqref{semiclassical-NS-eq}, has the following solution for the stationary state:
\begin{flalign}
u_x(y)=\frac{\rho a}{2\bar\eta} (y^2-yL_y).
\label{poiseuille-ux-eq}
\end{flalign}
Thus, we use this equation to calculate the numerical shear viscosity $\bar\eta$ of our models, which is needed to convert to physical units. The system size for the 2D model is $L_X\times L_Y = 256 \times 256$ with periodic boundary conditions in the $x$ direction and bounce-back in the $y$ direction and for the 3D model is $L_X\times L_Y \times L_Z = 256 \times 256 \times 1$ with periodic boundaries in the $x$ and $z$ directions and bounce back in the $y$ direction. An external electrical field of magnitude $E=10^{-8}$ in lattice units is is implemented as in Eq.\eqref{u-update-force-term-eq} ($a=E$ in natural units). Initially, we set $\mu =1.0$ and $\boldsymbol{u}=0$ everywhere for the two models. Note that the initial densities are different because we set equal $\mu$. In Fig. \ref{poiseuille-fig} we see the velocity profiles for five different relaxation times after $10^6$ time steps. A curve fit using Eq.\eqref{poiseuille-ux-eq} is made in the points given by the simulation, which provides the shear viscosity. As we can see in Fig. \ref{poiseuille-fig}, $\bar\eta$ the relation below is followed with good agreement by the two models:
\begin{flalign}
\bar\eta = \frac{1}{3}\left ( \tau - \frac{\Delta t}{2}  \right) .
\end{flalign}

\subsection{Ohm's law}\label{ohm-sec}

\begin{figure}[ht]
\center
\includegraphics[width=1.0\linewidth]{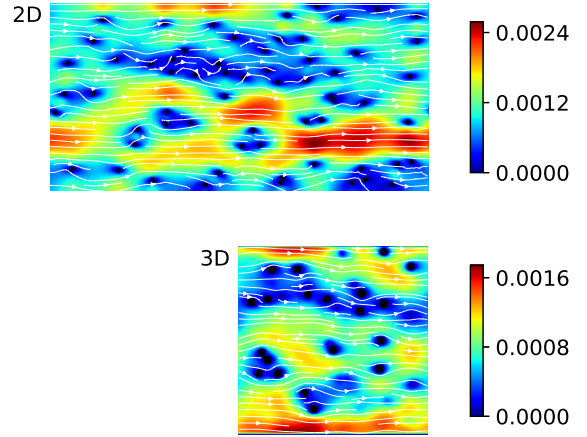}
\caption{(Color online) Steady state of the velocity field for the electrons passing through randomly placed obstacles to obtain the Ohm's law. The black objects are the obstacles, the colors represent the magnitude of the velocity field and the streamlines show its directions. The entire sample used in the 2D model is shown on the top while a cross section at $z=L_Z/2$ of the 3D sample can bee seem on the bottom. The electrical filed used was $E=10^{-7}$ in lattice units.}
\label{ohm-streamlines-fig}
\end{figure}

\begin{figure}[ht]
\center
\includegraphics[width=1.0\linewidth]{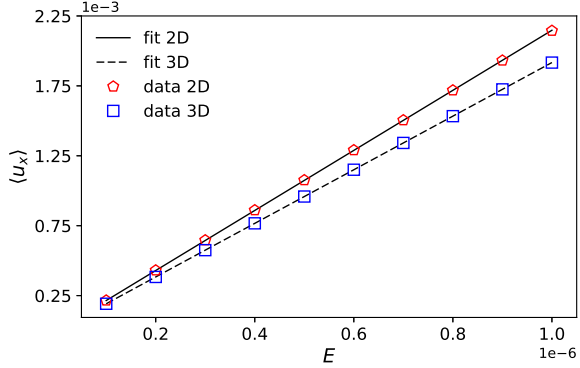}
\caption{(Color online) Linear relation between the average velocity in the $x$ direction and the external electrical field. This linear relation leads to the Ohm's law.}
\label{ohm-vel-fig}
\end{figure}

When an electrical current passes through an ohmic material (e.g, metals) they offer a resistance produced by the collisions between the electrons and a background of impurities and defects on the crystal lattice~\cite{ibach03}. The relation between the current $I$ and the applied electrical potential difference $V$ is linear: $V = R\, I$, where $R$ is the resistance. Here we model the electrical resistance with randomly placed obstacles through which the electrons flow. For the 2D sample with system size $L_X\times L_Y=512\times 256$, we sort 64 circles of radius 3 forming a porous medium with porosity $\phi_{2D} = 0.986$ while for the 3D sample, with system size $L_X\times L_Y \times L_Z = 128 \times 128 \times 128$, we sort 450 spheres of radius 3 forming a medium with porosity $\phi_{3D}=0.974$ (see Fig. \ref{ohm-streamlines-fig}). The relaxation time is set $\tau=0.9$. Initially, $\mu=1.0$ and $\boldsymbol{u}=0$ in the whole domain for the two models. We set periodic boundary conditions in the $x$ direction, slip-free conditions in the $y$ direction (also in the $z$ direction for the 3D model) and bounce-back conditions on the obstacle's surface. To measure the convergence with time, we calculate the relative error as the spatial average of $\vert |\boldsymbol{u}_{new}| - |\boldsymbol{u}_{old}| \vert / |\boldsymbol{u}_{new}|$ considering all fluid point with non-zero $\boldsymbol{u}_{new}$ and we stop the simulations when the error is smaller than $10^{-7}$. In Fig. \ref{ohm-vel-fig} we see that the average speed in the $x$ direction (considering all fluid points) have a linear relation with external electrical field $E$. This relation straightforwardly leads to the Ohm's law ($V=R\,I \Rightarrow I = \frac{L_X}{R}E$) since the current can be written as $I_{2D} = \rho  L_Y \phi \langle u_x \rangle$ for 2D and $I_{3D} = \rho  L_Y L_Z\phi \langle u_x \rangle$ for 3D (the densities are essentially constant in the whole domain for the steady state: $\rho_{2D} = 3.142$ and $\rho_{3D}=4.189$).

\section{Summary and conclusion}\label{conclusion-sec}

The two main goals of the present paper are the construction of the semiclassical LBM and the test of an isothermal LBM that simulates electrons in metals in the hydrodynamic regime. 

We have obtained an expansion of the equilibrium distribution function up to fourth order in generalized D-dimensional polynomials, orthonormal under a generic weight. The Hermite polynomials are just a particular case of these generalized polynomials where the weight is given by the gaussian function. The choice of a weight function close to the equilibrium distribution function renders  convergence attainable within a few terms in the truncated expansion. We extend here the concept of quadrature to the generic weight function of the polynomials thus beyong the Gauss-Hermite quadrature which is restricted to the Gauss-Hermite weight. In this way we generale the standard lattices used in the LBM of classical fluids to the semiclassical ones. The macroscopic equations for semiclassical fluids are obtained here through the Chapman-Enskog expansion. The notorious advantageous feature of the LBM is that the mass, momentum and energy conservation equations stem from a Chapman-Enskog expansion where the distribution function is expanded only up to first, third and fourth order~\cite{coelho14} in the orthonormal polynomials, respectively. This renders the same results as obtained using  the non-expanded distribution function. We show here that the forcing term for the semiclassical distribution satisfies the moment constraints up to second order even for the Lorentz force, which  depends on the microscopic velocity in case of the magnetic force.\\

An isothermal LBM for electrons in metals for two and three dimensions was developed using a weight near to the Fermi-Dirac  equilibrium distribution function. It is based on the expansion of the Fermi-Dirac distribution up to second order in generalized polynomials and uses the new D2V9 and D3V19 quadratures. We validate our model with the Riemann problem by comparing the density and velocity profiles of the shock waves with the analytical solution for the classical inviscid case. We also perform the Poiseuille flow, obtaining the expected parabolic profiles for the velocity. Lastly, we retrieved the Ohm's law by forcing the electrons to pass through a medium with randomly placed impurities (obstacles) analogously as a classical porous medium. We verified a linear relation between the applied external electrical field and the average velocity of the fluid in the steady state flow which leads to the Ohm's law.\\

The present semiclassical LBM opens the way for the modeling of  many other fluids such as made by bosons close to the Bose-Einstein condensation, as in Ref.~\cite{coelho2017fully}.  The semiclassical LBM allows for the investigation of the hydrodynamic limit of the electronic flow of many 2D novel materials, such as graphene~\cite{bandurin16}, topological insulators~\cite{PhysRevB.93.155122}, Weyl systems~\cite{Lucas23082016} and the 2D metal Palladium cobaltate~\cite{Moll1061}.

 \appendix


\section{Quadratures}\label{quadrature-appendix}

In Table \eqref{lattices-tab} we show more quadratures for the semiclassical LBM in 1D, 2D and 3D. The expressions for the D1V7 lattice can be found below.

\subsection{D1V7}

The lattice D1V7 can be used in models with higher orders EDF expansions sin it has order $M=9$. The geometrical velocities are $\{e_0, e_1,e_2,e_3,e_4,e_5,e_6\} = \{0, +1, -1, +2, -2, +3, -3\}$, and the weight are $w_0$ for $\alpha$=0, $w_1$ for $\alpha$ = 1 and 2, $w_2$ for $\alpha$ = 3 and 4 and $w_3$ for $\alpha$ = 5 and 6. Quadrature equations satisfied by the D1V7 lattice:
\begin{flalign*}
&\sum_\alpha w_\alpha = I_0 \Rightarrow w_0+2w_1+2w_2+2w_3 = I_0\\
&\sum_\alpha w_\alpha e_\alpha^2 = I_2c_s^{2} \Rightarrow 2w_1+8w_2+18w_3=I_2c_s^{2}\\
&\sum_\alpha w_\alpha e_\alpha^4 = 3I_4c_s^{4} \\ &\Rightarrow 2w_1+32w_2+162w_3=3I_4c_s^{4} \\
&\sum_\alpha w_\alpha e_\alpha^{6} = 15I_6 c_s^{6} \\ &\Rightarrow 2w_1+128w_2+1458w_3 = 15I_6c_s^{6}\\
&\sum_\alpha w_\alpha e_\alpha^8 = 105 I_8 c_s^{8}  \\&\Rightarrow 2w_1+512w_2+13122w_3 = 105 I_8c_s^{8}
\end{flalign*}
Solutions:
\begin{flalign*}
&w_0 = \frac{1}{360} (360I_0-150I_6c_s^{6} + 420 I_4 c_s^{4}-490I_2c_s^{2})\\
&w_1 = \frac{1}{16} (-13I_4c_s^{4} + 5 I_6 c_s^{6}+12I_2c_s^{2})\\
&w_2 = \frac{1}{120}(30I_4c_s^{4} - 15 I_6c_s^{6} - 9I_2c_s^{2})\\
&w_3 = \frac{1}{720} (15I_6c_s^{6}-15I_4c_s^{4}+ 4I_2c_s^{2})
\end{flalign*}
There are six solutions for $c_s$, which can be found by solving the equation
\begin{flalign*}
12I_2 - 49I_4c_s^{2} + 70I_6c_s^{4} - 35I_8 c_s^{6} =0.
\end{flalign*}
One of them is:
\begin{flalign*}
&c_s = \Big\{ \frac{2}{3} \frac{I_6}{I_8}- \frac{49\, 2^{1/3}I_4}{(B + \sqrt{4A^3+B^2})^{1/3}} \\ \nonumber
&+ \frac{140\, 2^{1/3}I_6^2 }{3I_8(B+ \sqrt{4A^3+B^2})^{1/3}} + \frac{(B + \sqrt{4A^3+ B^2})^{1/3}}{105 \, 2^{1/3}I_8} \Big\}^{1/2}
\end{flalign*}
where
\begin{flalign*}
&A = - 4900I_6^2 + 5145I_4I_8\\
&B = 686000I_6^3 - 1080450I_4I_6I_8 + 396900I_2I_8^2
\end{flalign*}

\begin{table*}[ht]
	\centering
\begin{tabular}{ |l |c| c|c| c |c| }
\hline
Lattice & $M$  &$\boldsymbol{e_\alpha}$ & p & $w_\alpha$ &  $c_s$\\ \hline \hline
D1V3 & 5 & $0$& 1& $I_0(1-J_2/3)$& $\sqrt{\frac{I_2}{3I_4}}$ \\
 & & $\pm 1$&2 & $I_0J_2/6$ & \\ \hline
D1V5 & 7 & $0$&1 & $I_0 - 10I_2c_s^2/9+I_4c_s^4/3$ & \\
($a$ and $b$)& & $\pm 1$&2 & $9I_2c_s^2/16 -3I_4c_s^4/16$&$  \sqrt{\frac{10I_4\pm \sqrt{100I_4^2-60I_6I_2}}{10I_6}}$  \\
 & & $\pm 3$&2 & $3I_4c_s^4/144-I_2c_s^2/144$& \\ \hline
D2V6 & 3 & $\left( \cos\frac{2\pi n}{6} ,\sin \frac{2\pi n}{6}  \right)$ & 6&$I_0/6$ & $\sqrt{\frac{I_0}{2I_2}}$ \\ \hline
D2V9 & 5 & $(0,0)$&1 & $I_0(1-5J_2/9)$ &  \\
 & & $(1,0)_{FS}$&4 & $I_0J_2/9$&  $\sqrt{\frac{I_2}{3I_4}}$ \\
 & & $(1,1)_{FS}$ & 4&$I_0J_2/36$ &  \\ \hline
 D3V15 & 5 & $(0,0,0)$&1 & $I_0(1-7J_2/9)$ &  \\
 & & $(1,0,0)_{FS}$&6 & $I_0J_2/9$&  $\sqrt{\frac{I_2}{3I_4}}$ \\
 & & $(1,1,1)_{FS}$ & 8&$I_0J_2/72$ &  \\ \hline
  D3V19 & 5 & $(0,0,0)$&1 & $I_0(1-2J_2/3)$ &  \\
 & & $(1,0,0)_{FS}$&6 & $I_0J_2/18$&  $\sqrt{\frac{I_2}{3I_4}}$ \\
 & & $(1,1,0)_{FS}$ & 12&$I_0J_2/36$ &  \\ \hline
D3V27 & 5 & $(0,0,0)$ & 1  & $I_0 - 2I_2^2/(3I_4)-I_6I_2^3/(27I_4^3)$ &  \\
 & & $(1,0,0)_{FS}$ & 6 & $(3I_2^2I_4^2+I_6I_2^3)/(54I_4^3)$ &  $\sqrt{\frac{I_2}{3I_4}}$\\
 & & $(1,1,0)_{FS}$ & 12  & $(3I_4^2I_2^2-I_6I_2^3)/(108I_4^3)$ &    \\
 & & $(1,1,1)_{FS}$ &  8  & $I_2^3I_6/(216I_4^3) $ & \\ \hline
\end{tabular}
\vspace{15pt}
\caption{Generalized lattices and their weights. $M$ is order of the quadrature (see Sec. \ref{quadrature-sec}) and p is the number of velocities with the same weight. The subscript $FS$ denotes a fully symmetric set of points. }
\label{lattices-tab}
\end{table*}

\section*{Acknowledgements}

R.C.V. Coelho thanks FAPERJ and the European Research Council (ERC) Advanced Grant 319968-FlowCCS for the financial support.

\bibliographystyle{elsarticle-num}
\bibliography{references}


\end{document}